\documentclass[12pt]{article}
\usepackage{latexsym}
\usepackage{epsfig,amssymb,euscript,slashed}
\usepackage[linktocpage]{hyperref}
\usepackage{amsmath}
\usepackage{cite} 
\usepackage{array,calc,epsfig}
\usepackage{soul}
\usepackage{bbm}
\usepackage{fancybox}

\usepackage{physics}

\usepackage{comment}
\usepackage{mathrsfs}

\usepackage{tikz}
\usepackage{lmodern}
\usepackage{subfig}
\usepackage{xcolor}
\usetikzlibrary{decorations.pathmorphing}
\usetikzlibrary{patterns}
\usetikzlibrary{shapes.misc}
\usetikzlibrary{decorations.pathreplacing}

\tikzset{
    photon/.style={decorate, decoration={snake}},
    electron/.style={postaction={decorate},
        decoration={markings,mark=at position .55 with {\arrow[draw=blue]{>}}}},
    gluon/.style={decorate,
        decoration={coil,amplitude=4pt, segment length=5pt}} 
}

\colorlet{myPurple}{blue!40!red}
\definecolor{Fiorentina}{RGB}{134, 41, 198}
\definecolor{DarkGreen}{RGB}{34,139,34}

\numberwithin{equation}{section} %%
\usepackage[]{graphicx}
\usepackage{color}

\newcommand{\be}{\begin{equation}}
\newcommand{\ee}{\end{equation}}
\newcommand{\beq}{\begin{equation}}

\newcommand{\eeq}{\end{equation}}

\newcommand{\half}{\frac{1}{2}}

\newcommand{\bs}{\begin{split}}
\newcommand{\es}{\end{split}}
\newcommand{\area}{\mathcal{A}}

\newcommand{\D}{\mathcal{D}}

\newcommand{\F}{\mathcal{F}}

\newcommand{\pd}{\partial}

\newcommand{\ep}{\varepsilon}

\begin{document}
\font\cmss=cmss10 \font\cmsss=cmss10 at 7pt

\begin{flushright}{
%\scriptsize DFPD-17-TH-xx \\  
}
\end{flushright}
\hfill
\vspace{18pt}
\begin{center}
{\Large 
\textbf{Holographic entanglement entropy and complexity of microstate geometries}
}

\end{center}

\vspace{8pt}
\begin{center}
{\textsl{Alessandro Bombini$^{\,a,b}$, Giulia Fardelli$^{\,c}$}}

\vspace{1cm}

\textit{\small ${}^a$ Department of Physics, Stockholm University, AlbaNova, 106 91 Stockholm, Sweden} \\  \vspace{6pt}

\textit{\small ${}^b$ Institut de Physique Th\'eorique, CEA Saclay, CNRS, 91191 Gif-sur-Yvette, France}\\

%\textit{\small ${}^a$ Dipartimento di Fisica ed Astronomia ``Galileo Galilei",  Universit\`a di Padova,\\Via Marzolo 8, 35131 Padova, Italy} \\  \vspace{6pt}

%\textit{\small ${}^b$ I.N.F.N. Sezione di Padova, Via Marzolo 8, 35131 Padova, Italy}\\
\vspace{6pt}

\textit{\small ${}^c$ Department of Physics and Astronomy, Uppsala University, \\ Box 516, SE-751 20 Uppsala, Sweden}\\ 

\vspace{6pt}

\end{center}

\vspace{12pt}

\begin{center}
\textbf{Abstract}
\end{center}

\vspace{4pt} {\small
\noindent 
We study  holographic entanglement entropy and  holographic complexity in a two-charge, $\frac{1}{4}$-BPS family of solutions of type IIB supergravity, controlled by one dimensionless parameter. All the geometries in this family are asymptotically AdS$_3 \times \mathbb{S}^3 \times \mathbb{T}^4$ and, varying the parameter that controls them, they interpolates between the global AdS$_3 \times \mathbb{S}^3 \times \mathbb{T}^4$ and the massless BTZ$_3 \times \mathbb{S}^3 \times \mathbb{T}^4$ geometry. Due to AdS/CFT duality, these geometries are dual to pure CFT heavy states.

We find that there is no emergence of entanglement shadow for all the values of the parameter and we discuss the relation with the massless BTZ result, underlying the relevance of the nature of the dual states. 

We also compute the holographic complexity of formation of these geometries, finding a nice monotonic function that interpolates between the pure AdS$_3$ result and the massless BTZ one.  
}

\vspace{1cm}

\thispagestyle{empty}

\vfill
\vskip 5.mm
\hrule width 5.cm
\vskip 2.mm
{
\noindent  {\scriptsize e-mails:  {\tt alessandro.bombini@fysik.su.se, giulia.fardelli@physics.uu.se} }
}

\setcounter{footnote}{0}
\setcounter{page}{0}

\newpage

%%%%%%%%%%%%%%%%%%%%%%%%%%%%%%%%%%%%%%%%%

\tableofcontents

\newpage
\section{Introduction}

The study of information-theoretical quantities such as entanglement entropy \cite{Calabrese:2004eu} and complexity \cite{Susskind:2014rva, Susskind:2014moa, Stanford:2014jda} is one of the relevant topics in the context of holography, due to the existence of various proposals for their dual interpretation \cite{Ryu:2006bv, Ryu:2006ef, Hubeny:2007xt, Alishahiha:2015rta, Brown:2015bva, Brown:2015lvg, Lehner:2016vdi, Chapman:2016hwi}. 
{In particular, the entanglement entropy, which is computable in CFT  through the ``replica trick" \cite{Calabrese:2009qy, Rangamani:2016dms}, can be determined holographically via the Ryu-Takayanagi prescription \cite{Ryu:2006bv, Ryu:2006ef}. According to this, the entanglement entropy of a given sub-region is computed as the area of the minimal surface, which enters in the bulk and is attached to its boundary\footnote{A nice computation on microstate geometries using relative entropy can be found in \cite{Michel:2018yta}.}.
%The entanglement entropy of a sub-region, that is computable in the CFT via the so-called replica trick \cite{Calabrese:2009qy, Rangamani:2016dms}, is holographically dual to the area of the minimal surface entering in the bulk that is attached to the boundary of such sub-region via the Ryu-Takayanagi prescription \cite{Ryu:2006bv, Ryu:2006ef}; 
The complexity, instead, has two different holographic proposals: the ``complexity=volume'' \cite{Susskind:2014rva, Stanford:2014jda}, that associates the complexity of the state to the volume of the co-dimension 1 maximal space-like manifold, and the ``complexity=action'' that associates it to the gravitational on-shell action computed on the so-called Wheeler-de Witt patch, taking into account all the possible boundary and corner terms \cite{Brown:2015bva, Brown:2015lvg, Lehner:2016vdi}. 

Our goal is to compute holographically both entanglement entropy and complexity in a family of type IIB supergravity solutions that are asymptotically AdS$_3 \times \mathbb{S}^3 \times \mathbb{T}^4$ and dual to pure heavy CFT states. Those states are element of the Hilbert space of the so-called D1D5 CFT \cite{Seiberg:1999xz, David:2002wn, Avery:2010qw}, that is a (1+1)-dimensional superconformal field theory with an $SU(2)_L \times SU(2)_R$ Kac-Moody algebra and an $SU(2)_1 \times SU(2)_2$ ``custodial'' global symmetry. This theory admits a special point in its moduli space, dubbed free-orbifold point, where the theory can be described by a (1+1)-dimensional non-linear sigma model whose target space is the symmetrized orbifold $(\mathbb{T}^4)^N/S_N$. In this work, we will specify to a particular one-parameter family of solutions of this kind, that are factorisable - i.e.~whose AdS$_3$ Einstein metric is independent of the coordinates of the $\mathbb{S}^3$ - and that interpolate between the vacuum global AdS$_3$ geometry and the BTZ black hole \cite{Balasubramanian:2005qu, Bombini:2017sge}.

After that, for these factorisable geometries in the de Donder gauge, we will prove that, in order to compute the entropy, one can reduce the computation of extremal co-dimensional 2 space-like surfaces on AdS$\times \mathbb{S}$ to a pure AdS lower-dimensional computation. We will put forward explicitly the computations of holographic entanglement entropy in a two-charge, $\frac{1}{4}$-BPS one-parameter family of solutions that are dual to well defined pure heavy states of the dual CFT theory; in this explicit case we will show that no entanglement shadow emerges \cite{Hubeny:2013gta, Freivogel:2014lja, Balasubramanian:2014sra}. We will then move to the computation of the complexity of such states via the ``complexity=volume'' proposal. 

The plan of the paper is the following: in sec.~\ref{sec:kmngeom}, we introduce the type IIB solutions,  setting up the system we are interested in, and then we discuss their holographic interpretation in terms of their dual heavy states. In sec.~\ref{sec:HEE6D} we prove how, for factorisable geometries in de Donder gauge, it is possible to reduce the computation to a pure AdS$_3$ problem. With this simplification, in sec.~\ref{sec:computingHEE} we describe the computations of the holographic entanglement entropy, {pointing out} % and we show
that no entanglement shadow emerges and {briefly} discussing the relation with known results for the BTZ black hole. Finally, in sec.~\ref{sec:complexity=volume}, we compute the complexity for these states. We close the paper with a discussion in sec.~\ref{sec:discussion}.

\section{Microstate Geometries}\label{sec:kmngeom}
In the context of type IIB string theory on ${\cal M}^{4,1} \times \mathbb{S}^1 \times \mathbb{T}^4$ a wide set of microstate geometries has been constructed, both of $\frac{1}{4}$-BPS and $\frac{1}{8}$-BPS nature \cite{Lunin:2001jy, Lunin:2002iz, Mathur:2003hj, Lunin:2004uu, Giusto:2004ip, Giusto:2004id, Skenderis:2006ah, Kanitscheider:2006zf, Skenderis:2007yb, Kanitscheider:2007wq, Skenderis:2008qn, Mathur:2011gz, Mathur:2012tj, Lunin:2012gp, Giusto:2013rxa, Giusto:2013bda, Bena:2011dd, Bena:2015bea, Bena:2016ypk, Bena:2017xbt, Bombini:2017got, Bakhshaei:2018vux, Bena:2017fvm, Bena:2018bbd, Heidmann:2018mtx, Walker:2019ntz, Heidmann:2019zws}. These are smooth, horizonless and asymptotically flat geometries, whose conserved charges are the same as the ``na\"ive'' D1D5P black hole and that, {in the \emph{decoupling region}, behaves approximately as AdS$_3\times \mathbb{S}^3 \times \mathbb{T}^4$}.
{These solutions can be} described by some scalar ``profile'' functions $Z_I$ (as well as by a set of 1- and 2-forms, that we will briefly describe later), similarly of what happens for the D1D5P black hole. The core difference is that for microstate geometries there are no singularities in the geometry nor poles in the profiles; moreover, due to supersymmetry \cite{Giusto:2013rxa}, the Asymptotically flat region is attached to the AdS$_3\times \mathbb{S}^3 \times \mathbb{T}^4$ region simply by ``adding back the 1'' in the profile functions $Z_1$, $Z_2$, that encodes the D1 and D5 charges \cite{Bena:2015bea, Bena:2016ypk, Bakhshaei:2018vux}.  {Finally}, we will always be able to work in the decoupling limit whose asymptotics is AdS$_3\times \mathbb{S}^3 \times \mathbb{T}^4$, that is well suited for the study of Holographic Entanglement Entropy \cite{Giusto:2014aba, Giusto:2015dfa, Moscato:2017usq}. 

{In this paper,} we will focus on recently build superstrata $(k,m,n)$ \cite{Giusto:2013rxa, Bena:2015bea, Bena:2016ypk, Bena:2017xbt, Bakhshaei:2018vux}: these solutions are controlled by three integers and they are invariant under rotation of the compact $\mathbb{T}^4$. The generic ansatz takes the factorised form $\mathbb{R}^{1,1} \times {\cal B}_4 \times \mathbb{T}^4$ 
\be\label{Skenderis}
\begin{split}
\dd s_{10}^2 &=\sqrt{ \frac{Z_1 Z_2}{\cal P}} \, \dd s_6^2 + \sqrt{\frac{Z_1}{Z_2}}\,  \dd s_{\mathbb{T}^4}^2, \quad \dd s_{\mathbb{T}^4}^2 = \sum_{i=1}^4 \dd z_i^2 \,, \\
e^{2\phi} &= \frac{Z_1^2}{\cal P}\,, \quad C_0 = \frac{Z_4}{Z_1} \,, \quad B_2 = \bar{B}_2  \,, \quad C_2 = \bar{C}_2 \,, \\
C_4 &= \bar{C}_4 + \frac{Z_4}{Z_2} \, \dd z^1 \wedge \dd z^2 \wedge \dd z^3 \wedge \dd z^4 \,, 
\end{split}
\ee 
where everything is $z_i$-independent. We have marked with an over-bar the forms which have legs only in the six-dimensional non-compact space, while we have explicitly written down the $\mathbb{T}^4$ directions.
These geometries can either be $\frac{1}{4}$- or $\frac{1}{8}$-BPS.  Restricting to the $v-$independent base ${\cal B}_4 = \mathbb{R}^4$, eq.~\eqref{Skenderis} becomes 
\be
\begin{split}
\dd s_6^2 &= - \frac{2}{\sqrt{\cal P}} \, (\dd  v  + \beta) \left[  \dd u + \omega + \frac{\cal F}{2} ( \dd  v  + \beta) \right] + \sqrt{\cal P} \ \dd s_4^2 \,, \\
\dd s_4^2 &= \Sigma \left( \frac{\dd r^2}{r^2+a^2} + \dd \theta^2\right) + (r^2+a^2) \sin^2 \theta \, \dd  \phi^2 + r^2 \cos^2 \theta \, \dd \psi^2 \,, \\
\Sigma &= r^2 +a^2 \cos^2 \theta\,, \quad {\cal P} = Z_1 Z_2 - Z_4^2 \,, \quad u = \frac{t-y}{\sqrt{2} } \,, \quad v = \frac{t+y}{\sqrt{2} }  \\
%
%\beta &=  \frac{R\, a^2}{\sqrt{2}\,\Sigma} \left(\sin^2 \theta \, \dd \phi - \cos^2 \theta \, \dd \psi\right)  \,, \quad \omega = \frac{R\, a^2}{\sqrt{2}\,\Sigma} \left(\sin^2 \theta \, \dd \phi + \cos^2 \theta \, \dd \psi\right)  \,, \\
%%
\bar B_2 &= - \frac{Z_4}{\cal P} \, ( \dd u + \omega) \wedge (\dd  v  + \beta) + a_4 \wedge (\dd  v  + \beta) + \delta_2 \,, \\
\bar C_2 &= - \frac{Z_2}{\cal P} \, ( \dd u + \omega) \wedge (\dd  v  + \beta) + a_1 \wedge (\dd  v  + \beta) + \gamma_2 \,, \\
%
%{\cal C}_2 &= - \frac{Z_1}{\cal P} \, ( du + \omega) \wedge (d v  + \beta) + a_2 \wedge (d v  + \beta) + \gamma_1  \,,\\
%
\bar C_4 &=  - \frac{Z_4}{\cal P} \, \gamma_2 \wedge   ( \dd u + \omega) \wedge (\dd  v  + \beta) + x_3 \wedge (\dd  v  + \beta) \,,
\end{split}
\ee
{When ${\cal F}= 0$ the geometry has two charges and it is $\frac{1}{4}$-BPS, while it has three charges and it is $\frac{1}{8}$-BPS otherwise.} {Notice that $\dd s_6^2$ is intended in the Einstein frame}. {It is worthy introducing some additional}  objects, that are gauge invariant under the remaining gauge freedom $B_2 \to B_2 + \dd  \lambda_1$, where $\lambda_1$ is an $u\,,v-$independent 1-form and have legs only on the base space $\mathbb{R}^4$~\cite{Bombini:2017got, Bakhshaei:2018vux}:
\be
\begin{split}
\Theta_1 \equiv \D a_1+ \dot \gamma_2 \,, \quad \Theta_2 \equiv \D a_2 + \dot \gamma_1 \,, \quad \Theta_4 \equiv \D a_4 + \dot \delta_2 \,,
\end{split}
\ee
where $\dot f= \pd_v f$ and where
\be
\D \equiv \dd_4 - \beta \wedge \pd_v \,.
\ee
{The functions introduced before should fulfil some constraints encoded in a set of differential equations, often informally dubbed ``layers''. The first layer is:}
%The objects that define the solution have to satisfy a set of equations, often informally dubbed ``layers'', that are linear partial differential equations if solved in order; the first layer is

\be\label{eq:Layer1old}
\begin{split}
*_4 \D \dot Z_1 & = \D \Theta_2 \,, \quad \D *_4 \D Z_1 = - \Theta_2 \wedge \dd \beta \,, \quad \Theta_2 = *_4 \Theta_2 \,, \\
*_4 \D \dot Z_2 & = \D \Theta_1 \,, \quad \D *_4 \D Z_2 = - \Theta_1 \wedge \dd \beta \,, \quad \Theta_1 = *_4 \Theta_1 \,, \\
*_4 \D \dot Z_4 & = \D \Theta_4 \,, \quad \D *_4 \D Z_4 = - \Theta_4 \wedge \dd \beta \,, \quad \Theta_4 = *_4 \Theta_4 \,, \\
\end{split}
\ee
while the second one is 
\be\label{eq:Layer2old}
\begin{split}
\D \omega + *_4 \D \omega_4 + {\cal F} \, \dd \beta &= Z_1 \Theta_1 + Z_2 \Theta_2 - 2 Z_4 \Theta_4 \,, \\
*_4 \D *_4 \left( \dot \omega - \half \D {\cal F} \right) &= \pd_v^2 (Z_1 Z_2 - Z_4^2)  - [\dot Z_1 \dot Z_2 - (\dot Z_4)^2] \\
& \qquad - \half *_4 (\Theta_1 \wedge \Theta_2 - \Theta_4 \wedge \Theta_4 ) .
\end{split}
\ee

The crucial point to be stressed is that these geometries have no causal-disconnecting horizon nor curvature singularities, they are geodesically complete and present the same conserved charges as the D1D5P black hole.
{Interpreted from the AdS$_3$/CFT$_2$ view point,} they are dual to heavy states in the CFT, which are \emph{pure states}, in contrast with the putative dual state of the na\"ive black hole geometry, which instead is a thermal one. 

\subsection{Factorisable geometries}

As noted in \cite{Bena:2017upb, Bena:2017xbt, Bena:2017fvm, Bombini:2017got}, there exists a class of superstrata, described in the previous section, that have a factorisable form. Using the notation of \cite{Bena:2016ypk}, we will focus on the case with $(k,m,n)=(1,0,n)$, whose metric is 
\be
\dd s_6^2 = - \frac{2}{\sqrt{\mathcal{P}}} (\dd v+\beta) \left[ \dd u+\omega +   \frac{\F}{2} (\dd v+\beta) \right] + \sqrt{\mathcal{P}} \, \dd s_4^2 \, ,
\ee
where we have
\begin{subequations}
\begin{align}
\hat v_{1,0,n} &= \frac{\sqrt{2}}{R} \, n v + \phi\,,\quad  \hat v_{2,0,2n} = \frac{\sqrt{2}}{R} \,2 n v +2 \phi\, ,  \\
\Delta_{1,0,n} &= \frac{a \, r^n}{(r^2+a^2)^{\frac{n+1}{2}}}  \, \sin \theta \, , \quad \Delta_{2,0,2n} =  \frac{a^2 \, r^{2n}}{(r^2+a^2)^{n+1}}  \, \sin^2 \theta\,, \\
Z_1 &= \frac{Q_1}{\Sigma} + \frac{R^2}{2 Q_5} \, b^2 \, \frac{\Delta_{2,0,2n}}{\Sigma} \cos \hat v_{2,0,2n} \, \quad Z_2 = \frac{Q_2}{\Sigma}\, , \quad Z_4 = b R \, \frac{\Delta_{1,0,n}}{\Sigma}\cos \hat v_{1,0,n} \,, \\
\omega &= \omega_0 + \frac{b^2}{a^2} \, \frac{a^2 R}{\sqrt{2}\,\Sigma} \left[ 1- \frac{r^{2n}}{(r^2+a^2)^n}   \right] \sin^2 \theta \, \dd \phi \,,\\
\beta &= \frac{a^2 R}{\sqrt{2}\,\Sigma} \left( \sin^2 \theta \, \dd \phi - \cos^2 \theta \, \dd \psi \right) \, ,\quad \omega_0 = \frac{a^2 R}{\sqrt{2}\,\Sigma} \left( \sin^2 \theta \, \dd \phi + \cos^2 \theta \, \dd \psi \right) ,\\
 \F &= -\frac{b^2}{a^2} \left[ 1- \frac{r^{2n}}{(r^2+a^2)^n} \right] \, ,  
\end{align}
\end{subequations}
so that, calling 
\be F_n \equiv  \left[ 1- \frac{r^{2n}}{(r^2+a^2)^n} \right] , 
\ee
we can write
\be
\omega =  \frac{a^2 R}{\sqrt{2}\,\Sigma} \left[ \left( 1+ \frac{b^2}{a^2} F_n  \right) \sin^2 \theta \, \dd \phi + \cos^2 \theta \, \dd \psi \right]  \, .
\ee

When $n\neq 0$ these are three-charge geometries, as can be easily seen from $F_n\neq 0$, which signals a non-vanishing momentum along the $\mathbb{S}^1$. 

We can write the full six-dimensional metric in the Einstein frame in the factorised form as a three-dimensional Asymptotically AdS$_3$ fibered along the $\mathbb{S}^3$,
\begin{equation}
    \dd s_6^2 = V^{-2} \tilde{g}_{\mu\nu} \dd x^\mu \dd x^\nu + G_{ab} (\dd \theta^a + A^a_\mu \dd x^\mu)(\dd \theta^b + A^b_\nu \dd x^\nu) , \quad V^{2} = \frac{\det G_{ab}}{(Q_1 Q_5)^{3/2}\sin^2 \theta \cos^2 \theta} \,,
\end{equation}
 where we have split the coordinates as $x^M = (x^\mu , \theta^a)$ with $x^\mu = (\tau, \sigma, r)$ and $\theta^a = (\theta, \phi, \psi)$,  $G_{ab}$ is a deformation of the $\mathbb{S}^3$ written in Hopf coordinates. We have defined the three dimensional Einstein metric as $\dd s^2 = \tilde{g}_{\mu\nu}\dd x^\mu \dd x^\nu$, 
\be\label{eq_metricn}
\begin{split}
\dd s^2 &= - \left[ r^2 \left( 1-\frac{b^2}{2a^2} \, F_n \right) + \frac{a^4}{a_0^2}    \right] \dd \tau^2 + \frac{b^2}{a^2}\, F_n \, r^2 d\tau d\sigma + r^2 \left( 1+ \frac{b^2}{2a^2} \, F_n  \right) \dd \sigma^2  \\
& \quad \, +\frac{r^2+\frac{a^4}{a_0^2} \left(1 + \frac{b^2}{2a^2} \, F_n  \right)}{(r^2+a^2)^2} \dd r^2 \,,
\end{split}
\ee
endowed with the regularity condition 
\be
 a^2 + \frac{b^2}{2} =  a_0^2 \equiv \frac{\sqrt{Q_1 Q_5}}{R}  \,.
\ee
The physical interpretation of these parameters will be clear in sec.~\ref{subsec:HolographicInterpretation}, once we have introduced the dual CFT description.

\subsection{The Two charge solution}
A useful subset of geometries, the one we will mainly use, is the $(k,0,0)$ type \cite{Bombini:2017sge}. They are  two-charge solutions, whose metric is generally non-factorisable, except when $k=1$. In the factorisable $(k,m,n)=(1,0,0)$ case some holographic studies have already been performed, and we will discuss them in the following section.  For this particular configuration, the metric takes the simple form:
\be\label{k=1geo}
\begin{split}
\frac{\dd s^2}{\sqrt{Q_1 Q_5}} &=  - \frac{r^2 + \tilde{a}^2}{Q_1 Q_5} \,  \dd t^2 +  \frac{r^2 + \tilde{a}^2}{(r^2+a^2)^2} \, \dd r^2 + \frac{r^2}{Q_1 Q_5} \, \dd y^2 \,,
\end{split}
\ee
where 
\be
\tilde{a}^2 \equiv \frac{a^2}{a_0^2} \, a^2 =\frac{a^2}{a^2 + \frac{b^2}{2}} \, a^2  \equiv \eta \, a^2 \,.
\ee
We are mainly interested in this one-parameter family of solutions, controlled by the parameter $\eta \in (0,1)$, because it shows some important and peculiar features. In particular, we can identify two interesting regimes in the $\eta$-parameter space:
\begin{itemize}
\item $\eta \to 0$, or $a \ll b$: in this region the geometry approaches the massless limit of the BTZ black hole \cite{Balasubramanian:2005qu, Bombini:2017sge}; 
\item $\eta \to 1$, or $a \gg b$:  the geometry reduces to vacuum AdS in global coordinates. 
\end{itemize}
It is known that the former geometry presents Entanglement Shadows \cite{Hubeny:2013gta, Balasubramanian:2014sra, Freivogel:2014lja}, while the latter presents none. The main goal of this paper is to study if these Shadows could actually appear in the intermediate region where our solutions live. We will introduce this concept more deeply in sec.~\ref{sec:HEE6D}.

\subsection{Holographic interpretation} \label{subsec:HolographicInterpretation}

All the geometries described above are supergravity solutions with an AdS$_3 \times \mathbb{S}^3 \times \mathbb{T}^4$ decoupling region, hence they are dual to a CFT, often dubbed D1D5 CFT \cite{Seiberg:1999xz, David:2002wn, Avery:2010qw, Giusto:2015dfa, Moscato:2017usq}. In the moduli space there exists a special point, called ``free-orbifold point'', where this theory can be described as a (1+1)-dimensional supersymmetric non-linear sigma model whose target space is the symmetric orbifold of $(\mathbb{T}^4)^N/S_N$. It has an $SO(4)_{\mathbb{S}^3}\simeq SU(2)_L \times SU(2)_R$ affine algebra of currents, dual to the isometries of $\mathbb{S}^3$, as well as a global ``custodial'' symmetry $SO(4)_{\mathbb{T}^4}\simeq SU(2)_1 \times SU(2)_2$, dual to the  isometries of $\mathbb{T}^4$. We will denote spinorial indexes $\alpha, \dot \alpha$ for $ SU(2)_L \times SU(2)_R$ and $A, \dot A$ for $ SU(2)_1 \times SU(2)_2$ to label the CFT operators. The elementary field content is made by Left- and Right-Moving bosons and fermions
\begin{equation}
    \left( \pd X^{A \dot A} (z) , \psi^{\alpha \dot A} (z)  \right) \oplus \left( \bar \pd X^{A \dot A} (\bar z) , \widetilde \psi^{\dot \alpha \dot A} (\bar z) \right) .
\end{equation}
In the theory, it exists a set of twist operators  that joins together $k$-copies of elementary strings - or \textit{strands} - to create a single strand of winding $k$; we will have then an untwisted sector and a twisted one (for more details, we refer to \cite{Avery:2010qw, Moscato:2017usq}). From the elementary fields it is possible to build the generators of the superconformal algebra :
\begin{equation}
\begin{aligned}
\label{eq:OPE1}
T(z)&=\frac{1}{2}\sum_{r=1}^N\epsilon_{\dot{A}\dot{B}}\epsilon_{AB}\partial X_{(r)}^{A\dot{A}}\partial X_{(r)}^{B\dot{B}}+\frac{1}{2}\sum_{r=1}^N\epsilon_{\alpha\beta}\epsilon_{\dot{A}\dot{B}}\psi_{(r)}^{\alpha\dot{A}}\partial\psi_{(r)}^{\beta\dot{B}} \,, \\
G^{\alpha A}(z)&=\sum_{r=1}^N\psi_{(r)}^{\alpha\dot{A}}\partial X^{\dot{B}A}_{(r)}\epsilon_{\dot{A}\dot{B}} \,,\quad J^{a}(z)=\frac{1}{4}\sum_{r=1}^N\epsilon_{\dot{A}\dot{B}}\psi_{(r)}^{\alpha\dot{A}}\epsilon_{\alpha\beta}(\sigma^{*a})^{\beta}{}_{\gamma}\psi_{(r)}^{\gamma\dot{B}} \,.
\end{aligned}
\end{equation}
Expanding them in modes, 
\begin{equation}
\label{eq:exp}
\begin{aligned}
T(z)&=\sum_nL_nz^{-n-2} \quad \iff L_n=\oint\frac{dz}{2\pi i} \, T(z)z^{n+1} \,, \\
J^a(z)&=\sum_nJ^a_nz^{-n-1} \quad \iff J^a_n=\oint\frac{dz}{2\pi i} \, J^a(z)z^{n}  \,,  \\
G^{\alpha A}(z)&=\sum_nG^{\alpha A}_nz^{-n-\frac{3}{2}} \,\iff G^{\alpha A}_n=\oint\frac{dz}{2\pi i} \, G^{\alpha A}(z)z^{n+\half} \,,
\end{aligned}
\end{equation}
it is possible to select the generator of the global subalgebra. The theory splits into two sectors: the Ramond and the Neveu-Schwarz sector; in the latter there exists a single vacuum $|0\rangle_{\rm NS}$, while the first one has 16 $R$- and custodial-charged vacua, both in the twisted and untwisted sector, that are
\begin{equation}
  \label{eq:2ch}
  |\alpha\dot{\alpha}\rangle_k,\,|AB\rangle_k,\,|\alpha B\rangle_k,\,|A\dot{\alpha}\rangle_k \, .
\end{equation}
The more relevant for us will be the highest-weight vacuum state $|++\rangle_k$ and the custodial-singlet $|00\rangle_k = \ep^{AB} |AB\rangle_k$. With these two vacua at hand, we can build the heavy states dual to the $(k,m,n)$ geometries of sec.~\ref{sec:kmngeom}  by means of the action of the generators of the global subalgebra; in fact, we have that the dual CFT state of the supergravity solution is \cite{Bena:2015bea, Bena:2016ypk}
\begin{equation}\label{eq:generalkmnqsuperstrata}
     \psi_{\{N^{(++)},N_{k,m,n}^{(00)}\}}\equiv \left(\left|++\right\rangle_1\right)^{N^{(++)}}\prod_{k,m,n}\left(\frac{\left(J^+_{-1}\right)^m}{m!}\frac{\left(L_{-1}-J_{-1}^3\right)^n}{n!} |00\rangle_k\right)^{N^{(00)}_{k,m,n}} \, ,
\end{equation}
with the constraint 
\begin{equation}
    N^{(++)} + \sum_{k,m,n} k N_{k,m,n}^{(00)} = N \,,
\end{equation}
and whose pictorial representation is furnished in fig.~\ref{fig:strandstate}.  For example, the simplest state whose supergravity dual has the two-charge geometry \eqref{k=1geo} is
\begin{equation}\label{eq:2chargeCFTstate}
     \psi_{\{N^{(++)},N^{(00)}\}}\equiv \left(\left|++\right\rangle_1\right)^{N^{(++)}} \left(  |00\rangle_1\right)^{N^{(00)}} \, ,
\end{equation}
with the constraint $N^{(++)} + N^{(00)} = N$. 

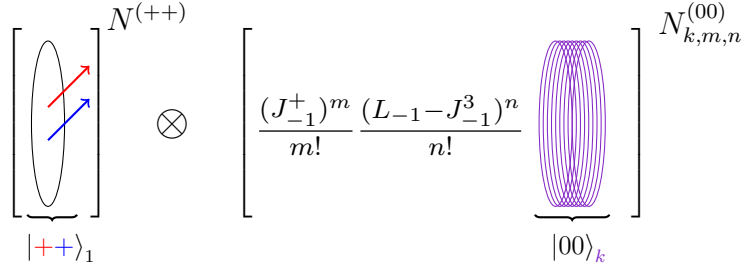
\begin{figure}
    \centering
    \subfloat{
    \begin{tikzpicture}[scale=1.1, baseline]
    \draw (0,0) ellipse (0.2 and 1);
    \draw [thick, red, ->]  (0,+0.2) -- (0.5, 0.7);
    \draw [thick, blue, ->]  (0,-0.2) -- (0.5, 0.3);
    \node at (-0.4,0) {$\left[  \vphantom{
    \begin{tikzpicture}[scale=1, baseline]
    \draw (0,0) ellipse (0.2 and 1);
    \draw [thick, red, ->]  (0,+0.2) -- (0.5, 0.7);
    \draw [thick, blue, ->]  (0,-0.2) -- (0.5, 0.3);
     \draw [thick, decorate, decoration={brace,amplitude=2pt, mirror}] (-0.25, -1.1) -- (0.25, -1.1);
    \end{tikzpicture}
    } \right.$};
    \node at (+0.6,0) {$\left.  \vphantom{
    \begin{tikzpicture}[scale=1, baseline]
    \draw (0,0) ellipse (0.2 and 1);
    \draw [thick, red, ->]  (0,+0.2) -- (0.5, 0.7);
    \draw [thick, blue, ->]  (0,-0.2) -- (0.5, 0.3);
     \draw [thick, decorate, decoration={brace,amplitude=2pt, mirror}] (-0.25, -1.1) -- (0.25, -1.1);
    \end{tikzpicture}
    } \right]$};
    \node at (1.2, 1.3) {$N^{(++)}$};
    \draw [thick, decorate, decoration={brace,amplitude=2pt, mirror}] (-0.25, -1.1) -- (0.35, -1.1);
    \node [scale=0.85] at (0, -1.5) {$ \phantom{-} \left| \textcolor{red}{+}\textcolor{blue}{+}\right\rangle_1$};
    \node [scale=1.4] at (1.5,0) {$\otimes$};
    \end{tikzpicture}
    }
    %%%%
    \subfloat{
    \begin{tikzpicture}[scale=1.1, baseline]
    \draw [Fiorentina] (0,0) ellipse (0.2 and 1);
    \draw [Fiorentina]  (0.05,0) ellipse (0.2 and 1);
    \draw [Fiorentina]  (0.10,0) ellipse (0.2 and 1);
    \draw [Fiorentina]  (0.15,0) ellipse (0.2 and 1);
    \draw [Fiorentina]  (0.20,0) ellipse (0.2 and 1);
    \draw [Fiorentina]  (0.25,0) ellipse (0.2 and 1);
    \draw [Fiorentina]  (0.30,0) ellipse (0.2 and 1);
    \draw [Fiorentina]  (0.35,0) ellipse (0.2 and 1);
    \draw [Fiorentina]  (0.40,0) ellipse (0.2 and 1);
    \node [scale=1.25] at (-2 ,0) {$\frac{(J_{-1}^+)^m}{m!} \frac{({L}_{-1}-J_{-1}^3)^n}{n!}$};
    \node at (-3.75,0) {$\left[  \vphantom{
    \begin{tikzpicture}[scale=1, baseline]
    \draw (0,0) ellipse (0.2 and 1);
    \draw [thick, red, ->]  (0,+0.2) -- (0.5, 0.7);
    \draw [thick, blue, ->]  (0,-0.2) -- (0.5, 0.3);
     \draw [thick, decorate, decoration={brace,amplitude=2pt, mirror}] (-0.25, -1.1) -- (0.25, -1.1);
    \end{tikzpicture}
    } \right.$};
    \node at (+1,0) {$\left.  \vphantom{
    \begin{tikzpicture}[scale=1, baseline]
    \draw (0,0) ellipse (0.2 and 1);
    \draw [thick, red, ->]  (0,+0.2) -- (0.5, 0.7);
    \draw [thick, blue, ->]  (0,-0.2) -- (0.5, 0.3);
    \draw [thick, decorate, decoration={brace,amplitude=2pt, mirror}] (-0.25, -1.1) -- (0.25, -1.1);
    \end{tikzpicture}
    } \right]$};
    \draw [thick, decorate, decoration={brace,amplitude=2pt, mirror}] (-0.25, -1.1) -- (0.65, -1.1);
    \node [scale=0.85] at (0, -1.5) {$ \phantom{--} \left| 00 \right\rangle_{\textcolor{Fiorentina}{k}}$};
    \node at (1.75, 1.2) {$N^{(00)}_{k,m,n}$};
    \end{tikzpicture}
    }
    \caption{A pictorial representation of the state \eqref{eq:generalkmnqsuperstrata}. We show the untwisted strands with \textcolor{red}{left}- and \textcolor{blue}{right}-$R$-charges $\left| \textcolor{red}{+}\textcolor{blue}{+}\right\rangle_1$ and the \textcolor{Fiorentina}{twisted} uncharged $|00\rangle_{\textcolor{Fiorentina}{k}}$ ones. }
    \label{fig:strandstate}
\end{figure}

The duality relates $N^{(00)}$, $N^{(++)}$ with the supergravity parameters $a$, $b$ controlling the geometry through the relations:
\begin{equation}
    N^{(++)} + N^{(00)} = N \leftrightarrow  a^2 + \frac{b^2}{2} = a_0^2 \,,
\end{equation}
and  precisely %the precise holography gives the identification
\begin{equation}
    \frac{N^{(++)}}{N} = \frac{a^2}{a_0^2} \equiv \eta \,, \quad \frac{N^{(00)}}{N} = \frac{b^2}{2a_0^2} \equiv \ep = 1- \eta \,,
\end{equation}
where we have introduced the two adimensional parameters $\eta= 1- \ep$ that are sometimes used in the literature. As said on the supergravity side, when $a\to a_0$ ($\eta\to 1$) the geometry approaches the vacuum AdS$_3 \times \mathbb{S}^3 \times \mathbb{T}^4$; this is easy to see in the CFT side since the state $|\!++\rangle_1^N$ goes under spectral flow to the NS-vacuum $|0\rangle_1^N$. On the contrary, when $a\to 0$ ($\eta \to 0$), on the supergravity side it approaches the massless BTZ geometry, while the state in the CFT approaches the pure state $|00\rangle_1^N$; this simply means that the geometry dual to the pure state $|00\rangle_1^N$ differs from the one of the thermal state dual to the na\"ive black hole geometry by  stringy corrections that are not captured in the supergravity regime under scrutiny. This fact will be relevant in the following discussion when we will stress the difference between the holographic results on pure (micro)states and on the na\"ive massless BTZ geometry dual to a thermal state. 

\subsection{The late behaviour and emergence of a ``quantum'' scale}

As briefly mentioned in the previous section, some investigation on the holography of the microstate geometries have been conducted; for example, a set of nice computations of 4-point functions involving the two heavy operators dual to the geometry and two light operators dual to supergravity mode on top on that geometry were computed \cite{Giusto:2015dfa, Galliani:2016cai, Galliani:2017jlg, Bombini:2017got, Bombini:2019vnc, Tian:2019ash, Bena:2019azk}. We will now focus on the results of \cite{Bombini:2017sge}, since there the computation on the geometry \eqref{k=1geo} was put forward. There the authors computed the HHLL correlator 
\begin{equation}
    {\cal C} (z, \bar z) = \langle \bar O_H (0) \bar O_H (\infty) O_L (1) \bar O_L (z,\bar z) \rangle ,
\end{equation}
involving two light operators made with the two elementary bosonic fields
\begin{equation}
     O_{\mathrm{Bos}} (z, \bar z) = \sum_{r=1}^N \frac{\ep_{\dot A \dot B}}{\sqrt{2N}} \, \pd X^{1 \dot A}_{(r)} (z) \bar \pd X^{1 \dot B}_{(r)} (\bar z)  , \quad \bar O_{\mathrm{Bos}} (z, \bar z) = \sum_{r=1}^N \frac{\ep_{\dot A \dot B}}{\sqrt{2N}} \, \pd X^{2 \dot A}_{(r)} (z) \bar \pd X^{2 \dot B}_{(r)} (\bar z) ,
\end{equation}
and two heavy operators \eqref{eq:2chargeCFTstate}, finding that\footnote{We recall that the plane/cylinder map is
$$z= e^{i (\tau + \sigma)}\,, \quad \bar z= e^{i(\tau - \sigma)} \,.$$}
\begin{equation}
     {\cal C} (\tau,  \sigma) = \,\frac{a}{a_0} (\pd_\tau^2 -\pd_\sigma^2) \sum_{\ell \in\mathbb{Z}} e^{i \ell \sigma}\sum_{n=1}^\infty \frac{\exp\left[-i\frac{a}{a_0}\sqrt{(|\ell| + 2 n)^2+ \frac{b^2  \ell^2}{2 a^2}}\tau\right]}{\sqrt{\ell + \frac{b^2}{2 a^2}\frac{\ell^2}{(|\ell|+2n)^2}} }\,. 
\end{equation}
In the $a\to 0$ limit it is easy to see that this correlator approaches
\begin{equation}
    {\cal C}_{\rm micro} (\tau, \sigma) \sim  (\pd_\tau^2 -\pd_\sigma^2) \left[\frac{a^2}{a_0^2} \frac{1}{1- e^{i \, \frac{a^2}{a_0^2} \, \tau} }  \left( \frac{1}{1-e^{i( \sigma+ \tau)}}   + \frac{1}{1-e^{i( \sigma - \tau )}}   -1 \right) \right] , 
\end{equation}
that has to be contrasted with the result on the na\"ive massless BTZ black hole 
\begin{equation}
    {\cal C}_{\rm BTZ} (\tau, \sigma) \sim (\pd_\tau^2 -\pd_\sigma^2) \left[ \frac{1}{\tau} \left( \frac{1}{1-e^{i( \sigma+ \tau)}}   + \frac{1}{1-e^{i( \sigma - \tau )}}   -1 \right) \right] ;
\end{equation}
It is easy to see that the two results agree up to a time scale $\tau \sim a_0^2/a^2$, that is when there is no time for the perturbation to probe the different geometric structure of the two geometries; after that, the correlator computed on the na\"ive geometry decays exponentially while the other one start oscillating indefinitely, as prescribed by unitarity. We report a pictorial representation of that in fig.~\ref{fig:pBTZ}.
\begin{figure}
    \centering
    \includegraphics[scale=0.55]{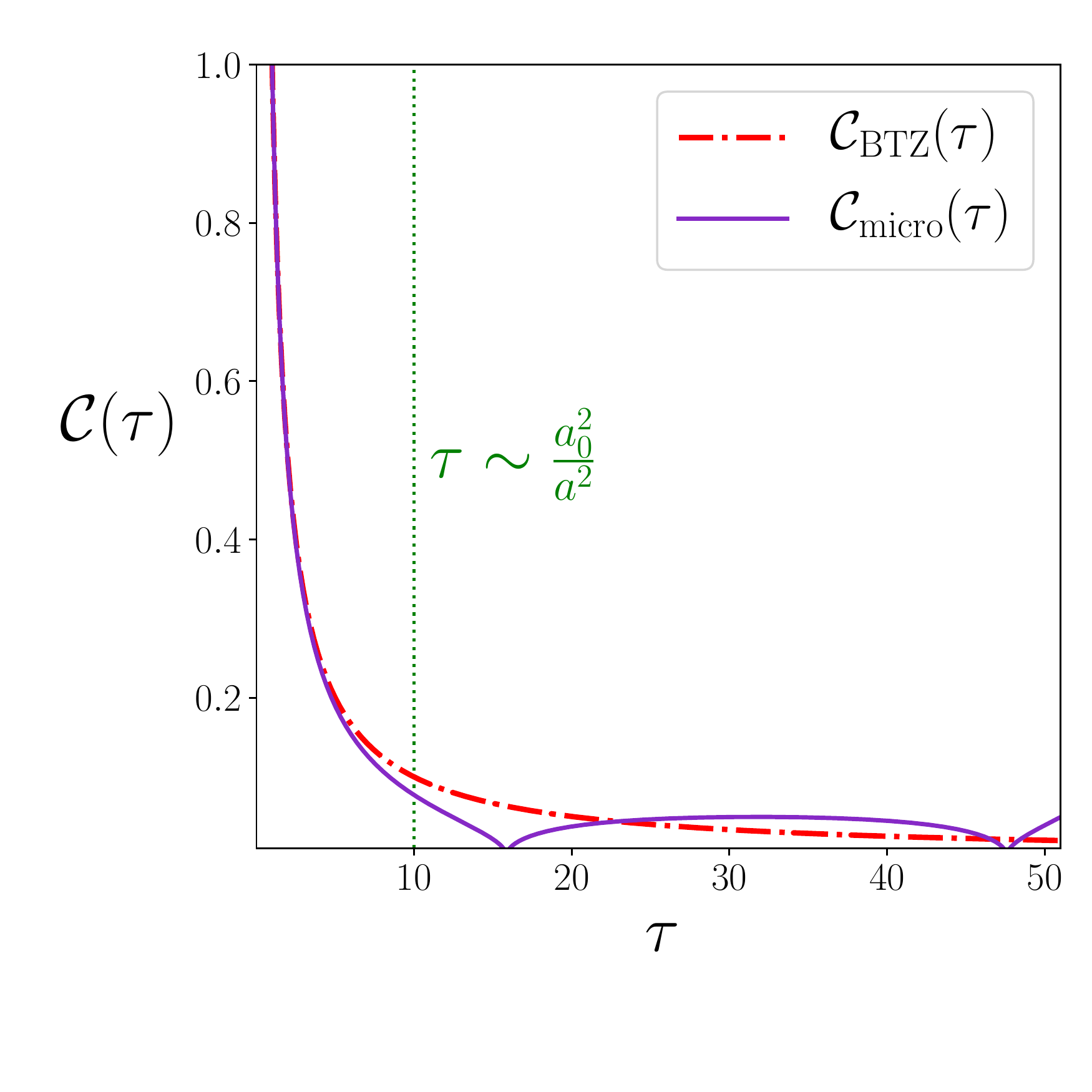}
    \caption{A pictorial representation of the HHLL correlator computed in both the na\"ive massless BTZ geometry (in dash-dotted red) and in the pure microstate geometry (in violet). Up to a certain time $\tau \sim a_0^2/a^2$, the two correlators present the same decaying behaviour; after that, the two starts to differ: the BTZ one maintains its decaying behaviour, while the microstate one starts oscillating, as imposed by unitarity. }
    \label{fig:pBTZ}
\end{figure}

We have thus seen appearing a dramatic difference of an interesting observable in the two geometries, that signals a different behaviour of observables between pure and thermal states. We want to study other interesting observables on microstates in order to understand better the (holographic) properties of these geometries. In the following sections, we will thus focus on the study of Holographic Entanglement Entropy and Holographic Complexity.

% ---------------------------------------------------------------------------------------------------------------------------------

\section{Holographic Entanglement Entropy in separable six-dimensional geometries}\label{sec:HEE6D}

One of the most relevant observables that can be computed holographically is the \emph{Holographic Entanglement Entropy} (HEE) via the so-called \emph{Ryu-Takayanagi} (RT) formula \cite{Ryu:2006bv, Hubeny:2007xt, Rangamani:2016dms, Nishioka:2009un, Nishioka:2018khk }. The main idea is that one can associate the Entanglement Entropy of a subregion ${\frak A}$ on the AdS-boundary, where the CFT lives, to the Area  $\mathscr{A}$ of the minimal area surface in the bulk, %of the extremal surface having minimal area in the bulk $\mathscr{A}$ 
whose boundary is the same as the one of ${\frak A}$, i.e.
\begin{equation}
    S_{\rm HEE} = \min_{\pd \mathscr{A} = \pd {\frak A}} \frac{{\cal A}[\mathscr{A}]}{4 G_N} \,.
\end{equation}
One of the issue that may arise in the study is the emergence of an \emph{Entanglement Plateux} \cite{Hubeny:2013gta} or \emph{entanglement shadow} \cite{Freivogel:2014lja, Balasubramanian:2014sra}; it consists in a region unexplored by any minimal surface attached to any possible region on the boundary {and that could lead to a limit to the possibility of a complete bulk reconstruction} %; it is associated to some limit on the possible bulk reconstruction 
 from CFT data. {It is known that this phenomenon appears } on black hole solutions, as well as in some higher-dimensional bubbling geometries, e.g.~on the so-called \emph{Lin-Lunin-Maldacena} (LLM) geometries \cite{Balasubramanian:2017hgy}.

We thus would like to address the computation of holographic entanglement entropy in microstate geometries\footnote{A program started in \cite{Giusto:2014aba, Giusto:2015dfa}.} for all the possible intervals on the boundary CFT, focusing on the study of possible entanglement shadows in the two-charge geometry \eqref{k=1geo}. As we have already said, this type IIB solution is a one-parameter family of geometries that is factorisable as AAdS$_3\times \mathbb{S}^3 \times \mathbb{T}^4$ and that somehow interpolates between the pure global vacuum AdS$_3\times \mathbb{S}^3 \times \mathbb{T}^4$ (when $a\to a_0$) and the massless BTZ$_3\times \mathbb{S}^3 \times \mathbb{T}^4$ geometry (when $a\to0$). Since the former shows no entanglement shadow while the latter does \cite{Freivogel:2014lja}\footnote{This is due to the fact that the shadow can be easily computed on the massive non-rotating BTZ geometry \cite{Freivogel:2014lja} and then, since the three-dimensional AdS gravity has a mass-gap \cite{Carlip:1995zj}, i.e.~the $M\to 0$ limit gives a black hole geometry with point-sized horizon, we retain in such limit a non-vanishing entanglement shadow whose size is related to the AdS radius.}, we are interested in addressing this issue for generic values of the ratio $a/a_0$. In order to compute it we will need some ingredients: 
\begin{itemize}
    \item A slight modification of the RT proposal for geometries that have only an AdS factor, a proposal that was put forward in \cite{Giusto:2014aba, Giusto:2015dfa};
    \item A proof that this proposal reduces to compute the HEE via standard RT on AdS$_3$ for factorizable (or separable) geometries, allowing us to reduce consistently to three dimensions. This proof will be stated in sec.~\ref{sec:reducingformula}.
    \item A numerical procedure for extracting the shadow radius and then for computing the entropy. 
\end{itemize}
After all these steps, we will show that for any value of the ratio $a/a_0$, the geometry \eqref{k=1geo} shows no entanglement shadow whatsoever, even for $a$ approaching zero. One thus may  wonder how the non-vanishing shadow result of \cite{Freivogel:2014lja} is obtained in the limit $a\to 0$. It is due to a somewhat non-trivial mechanism: what happens is that, at a certain value of the ratio $a/a_0$, namely $a/a_0 \sim 0.55$ corresponding to approximately to a dual CFT pure state that has as many $|00\rangle_1$ strands as $|\!++\rangle_1$ strands, the non-linear equation that fixes the shadow radius $r_*$ in terms of the size of the boundary interval $\alpha \in (0, \pi)$ starts to admit two possible solutions, an $r_*=0$ solution (valid $\forall a$) and an $r_* \neq 0$ one, that instead exists only for $a\in (0,0.55)a_0$. This means that for some values of the ratio $a/a_0$ there exist two extremal surfaces that are connected with the interval; the key point is that the one with $r_* \neq 0$ is maximal, while the other with $r_* =0$ in minimal (and thus its area computes the HEE).  

While this seems to hold for all possible values of the ratio $a/a_0$, the specific case $a=0$ should be treated with extreme care, since the $r=0$ point is the location of the point-size horizon and this it is \emph{not} part of the spacetime. This means that we are not allowed to keep the $r_* = 0$ geodesic since it becomes unphysical, leaving us with the $r_* \neq 0$ geodesic as the only extremal geodesic whose boundary is the same as the boundary of the interval, reproducing the result known in the literature \cite{Freivogel:2014lja}\footnote{In particular, setting $a=0$ at the beginning forbids us to find the root $r_*=0$, since the non-linear equation that relates $r_*$ with the interval size at the boundary $l/R$ degenerates and admits only one solution, the one with $r_*>0$. }.

\subsection{Ryu-Takayanagi formula in separable geometries and reducing formula}\label{sec:reducingformula}

One of the issues that arises in the study of entanglement entropy in microstate geometries is that these have a truly six-dimensional nature and they reduce to an AdS$_3 \times \mathbb{S}^3$ structure only asymptotically. It is thus necessary to extend the definition of the Ryu-Takayanagi formula suitably.  A proposal to extend the prescription for calculating the HEE was presented in \cite{Giusto:2014aba}; it applies to stationary geometry that have AdS$\times \mathbb{S}$ asymptotics,  and generalises the covariant formula put forward in \cite{Ryu:2006bv, Hubeny:2007xt}. According to \cite{Giusto:2014aba}, given a 1D spacial region $A=\left[0,l  \right]$, the HEE should be defined as:
\begin{align}
\label{eq:6dee}
S_A=\frac{\area[{\frak A}] }{4 G_N} \,,
\end{align} 
where $\mathfrak{A}$ is a co-dimensional 2 extremal surface in the six dimension spacetime such that at the boundary it reduces to $\partial A\times \mathbb{S}^3$, and where $G_N$ is the six-dimensional Newton constant. 

Here we prove that for a special class of geometries it is possible to reduce the six-dimensional problem to a three-dimensional one, to whom the original RT prescription applies.  

We have shown in sec.~\ref{sec:kmngeom} that it is always possible to rewrite our six-dimensional metric with an almost product structure as
\begin{align}
\label{metricfact}
\dd s^2_6 =  {g}_{\mu \nu} \dd x^{\mu} \dd x^{\nu}+ G_{a b}\left(\dd \theta^{a}+A_{\mu}^{a }\dd x^{\mu} \right)\left(\dd \theta^{b}+A_{\nu}^{b}\dd x^{\nu} \right) \,,
\end{align}
where  we have split the six dimensional coordinate system as $x^M = (x^\mu \,, \, \theta^a)$. Now  we  define 
\begin{align}
\tilde{g}_{\mu\nu}^E \equiv g_{\mu \nu} \frac{\det G}{\det G_0}
\end{align}
which is the  reduced three-dimensional Einstein metric, where $\det G=\det G_{\alpha\beta}$ and $G_0$ is the round metric of $\mathbb{S}^3$. 

For some 3-charge solutions, in particular for the superstrata $(1,0,n)$ in eq.~\eqref{eq:generalkmnqsuperstrata}, $\tilde{g}_{\mu\nu}^E$ does not depend on the coordinates on $\mathbb{S}^3$. For a metric that has this property, it is possible to prove that the extremal surface in six dimensions is equivalent to $x^{\mu}\times \mathbb{S}^3$, where $x^{\mu}(\lambda)$ is a geodesic on $\tilde{g}_{\mu\nu}^E$. This greatly simplifies the computation of the HEE since it allows to restrict our attention to the pure three-dimensional part of the metric, which is asymptotically AdS and whose HEE can be computed via the usual RT prescription:
\begin{align}
\label{enen3}
S_A=\frac{L_{\gamma}}{4 G_N}=\frac{c}{6}\frac{L_{\gamma}}{R_{\text{AdS}}}= n_1 n_5 \frac{L_{\gamma}}{R_{\text{AdS}}} \,,
\end{align} 
where $L_\gamma$ is the lentgh of the space-like geodesic with minimal area on the AdS$_3$ part of the metric. In addition we have used the fact that for microstate geometries we have 
\begin{align*}
G_N=\frac{3}{2}\frac{R_{\text{AdS}}}{c}, \qquad\qquad c=6 n_1 n_5  \,, 
\end{align*}
where $c$  is the central charge of the dual CFT, where $n_1$, $n_5$ are the numbers of D1, D5 branes and $R_{\text{AdS}}$ the AdS radius.

According to eq.~(\ref{eq:6dee}), in the full six-dimensional metric we need to find an extremal four-dimensional surface in order to compute the HEE, which we parametrize as $x^I\left(\lambda, \theta^a \right)$, $I=\left\lbrace \mu, a \right\rbrace$\footnote{A generic parametrization is $x^I(\lambda, \theta^{a})$. Since the area functional contains an integral over $\theta^{a}$, and is thus invariant under reparametrization of $\xi^{a}$, we can identify $\xi^{a}$ with the space-time coordinate $\theta^{a}$ ($\xi^{a}=\theta^{a}$) without loss of generality.}. The induced metric on the submanifold is then
\begin{align}
\label{eq:metricsubmanifold}
\dd s^2_*=g_{\mu \nu} \dd x^{\mu} _* \dd x^{\nu}_*+G_{ab}\left(\dd  \theta^{a}+A_{\mu}^{a} \dd x_*^{\mu} \right)\left(\dd  \theta^{b}+A_{\nu}^{b} \dd x_*^{\nu} \right)\equiv g^*_{I J} \dd x^I \dd x^J \,,
\end{align}
where we have defined 
\begin{align*}
\dd x_*^{\mu}=\dot{x}^{\mu}\dd \lambda+\partial_{a}x^{\mu} \dd \theta^{a} \,,
\end{align*}
and where we have denoted $\dot{x}^{\mu} \equiv \pdv{x^{\mu}}{\lambda}$.

Following the idea of  Ryu and Takayanagi, $x^I$ should extremize the area functional 
\begin{align}
\label{eq:areafunctional}
I\left[x^\mu(\lambda, \theta^{a})\right]=\int \dd ^3 \theta\, \dd \lambda \, \sqrt{\text{det}g_*} \,.
\end{align}

We want to find under which assumptions the minimization problem in higher dimensions is equivalent to the one in three dimensions. In terms of extremal surfaces, we want to show which conditions the original metric must satisfy in order for  $\bar{x}^I=\left( x^{\mu}(\lambda),\theta^{a} \right)$ (where $x^{\mu}(\lambda)$ is a geodesic of $\tilde{g}^E$) to be a solution of the minimization problem (\ref{eq:areafunctional}). Extremizing the functional $I$ is equivalent to solve Euler-Lagrange equations:
\begin{align}
\label{eq:EL6D}
&\frac{\delta \sqrt{\text{det}g_*}}{\delta x^{\rho}}- \frac{\pd}{\pd \lambda} \, \frac{\delta \sqrt{\det g_*}}{\delta \dot x^\rho} - \frac{\pd}{\pd \theta^a} \, \frac{\delta  \sqrt{\det g_*}}{\delta \pd_a x^\mu} =0
\end{align}
In order to prove that the three-dimensional geodesic is a solution, we will compute eq.~(\ref{eq:EL6D}) in $\bar{x}$ or equivalently considering $\partial_{a}x^{\mu}=0$.

Let us look in detail at the induced metric  (\ref{eq:metricsubmanifold})\footnote{For later use, we report the induced metric evaluated at the solution:
\begin{align}
\label{eq:inducedsolution}
g^* \big|_{\bar{x}^I} \equiv \bar{g}=\begin{pmatrix}
\left( g_{\mu \nu}+ A_{\mu}^{a}A_{\nu}^{b} G_{a b}\right)\dot{x}^{\mu}\dot{x}^{\nu} & G_{c a} A_{\mu}^{c} \dot{x}^{\mu} \\
G_{c a} A_{\mu}^{c} \dot{x}^{\mu} & G_{a b} 
\end{pmatrix}\,,
\end{align}
and its inverse:
\begin{align}
\label{eq:inverse}
\bar{g}^{\lambda\lambda}=g^{\lambda\lambda},\qquad \bar{g}^{\lambda a}=-g^{\lambda\lambda} A_{\mu}^{a}\dot{x}^{\mu},\qquad \bar{g}^{a b}=G^{a b}+g^{\lambda\lambda}A_{\mu}^{a}A_{\nu}^{b}\dot{x}^{\mu}\dot{x}^{\nu}  \,, 
\end{align}
where we have denoted $g^{\lambda\lambda}$ the inverse of:
\begin{align*}
g_{\lambda\lambda}\equiv g_{\mu\nu}\dot{x}^{\mu}\dot{x}^{\nu} \,.
\end{align*}
Moreover we have 
\begin{align}
\label{eq:determinant}
\det \bar{g}=g_{\mu\nu}\dot{x}^{\mu}\dot{x}^{\nu} \det \left(G_{\alpha\beta} \right)=\tilde{g}_{\mu \nu}^E \dot{x}^{\mu}\dot{x}^{\nu} \det\left( G_0\right) .
\end{align}
}:
\begin{align*}
g^*_{\lambda \lambda}&=\left( g_{\mu \nu}+ A_{\mu}^{a}A_{\nu}^{b} G_{a b}\right)\dot{x}^{\mu}\dot{x}^{\nu}  \,,\\
g^*_{\lambda a}&= \left(g_{\mu \nu}+G_{cd}A_{\mu}^{c}A_{\nu}^{d} \right) \partial_{a} x^{\mu} \dot{x}^{\nu}+G_{ac} A_{\mu}^{c}\dot{x}^{\mu}  \,, \\
g^*_{a b}&= G_{ab}+ \left( g_{\mu \nu}+ G_{cd}A_{\mu}^{c}A_{\nu}^{d} \right)\partial_{a} x^{\mu}\partial_{b} x^{\nu}+A_{\mu}^{c}\left( G_{ac}\partial_{b}x^{\mu}+G_{bc}\partial_{a}x^{\mu} \right) \,.
\end{align*} 

Looking at the components of $g^*_{I J}$ we immediately realise that, when considering derivatives w.r.t.~$x^\rho$ and $\dot{x}^{\rho}$ in (\ref{eq:EL6D}), there is no difference in differentiating the full induced metric or directly $\bar{g}$, since terms proportional to $\partial_{a} x^{\mu}$ are not involved in differentiation and can be simply put to zero. It then implies that 
\begin{align}
\frac{\delta \sqrt{\det g_*}}{\delta x^{\rho}} \Bigg |_{\bar{x}^I}& = \frac{\delta \sqrt{\det \bar{g}}}{\delta x^{\rho}}  \, \\
\pdv{}{\lambda}\frac{\delta \sqrt{\det g_*}}{\delta \dot{x}^{\rho}} \Bigg |_{\bar{x}^I}& = \pdv{}{\lambda}\frac{\delta \sqrt{\det \bar{g}}}{ \delta \dot{x}^{\rho}} \,. 
\end{align}
The only non trivial term in the Euler-Lagrange equations is the last one, but we can see that
\begin{align}
\left( \pdv{}{\theta^a} \frac{\delta \sqrt{\det g_*}}{ \delta \partial_{a}x^{\rho}} \right) \Bigg|_{\bar{x}^I}= \pdv{}{\theta^{a}}\left( \frac{\delta \sqrt{\det g_*}}{\delta \partial_{a}x^{\rho}} \Bigg|_{\bar{x}^I}\right)  \,.
\end{align}
Firstly, we have 
\begin{align}
\label{eq:derivative}
\frac{\delta \sqrt{\det g_*}}{\delta \partial_{a}x^{\rho}} \Bigg|_{\bar{x}^I}=\frac{\sqrt{\det \bar{g}}}{2} \bar{g}^{I J}\frac{\delta g^*_{I J}}{\delta \partial_{a}x^{\rho}}\Bigg|_{\bar{x}^I} \,,
\end{align}
where the only components that contribute are $( I ,J  )=(\lambda, a)$ or $(a, b)$,  and computed in the solution they give
\begin{align*}
\frac{\delta g^*_{\lambda a}}{\delta \partial_{b}x^{\rho}}=\left( g_{\mu \rho}+G_{c d}A_{\mu}^{c}A_{\rho}^{d} \right)\dot{x}^{\mu}\delta_{a}^{b} \,, \\
\frac{\delta g^*_{ a b}}{\delta \partial_{d}x^{\rho}} = \left( G_{a c}\delta_{b}^{d}+G_{b c}\delta_{a}^{d} \right)  A_{\rho}^{c}  \,,
\end{align*}
so that
\begin{align}
\frac{\delta \sqrt{\det g_*}}{\delta \partial_{a}x^{\rho}} \Bigg|_{\bar{x}^I}=\sqrt{\det \bar{g}}\left( A_{\rho}^{a}-g^{\lambda\lambda}g_{\mu \rho} A_{\nu}^{a}\dot{x}^{\mu}\dot{x}^{\nu} \right) .
\end{align}
We can finally perform the last derivative, substituting $g_{\mu\nu}$ with $\tilde{g}_{\mu\nu}^E$ and defining 
\be
g^{\lambda\lambda}=\frac{1}{g_{\lambda\lambda}}=\frac{1}{\tilde{g}_{\mu\nu}^E\dot{x}^{\mu}\dot{x}^{\nu}}\cdot\frac{\det (G_0)}{\det G}\equiv \tilde{g}^{\lambda\lambda}\frac{\det (G_0)}{\det G},
\ee
finding that
\begin{equation}
\begin{split}
 \label{eq:xiderivative}
\pdv{}{\theta^{a}}\frac{\delta \sqrt{\det g_*}}{\delta \partial_{a}x^{\rho}} \Bigg|_{\bar{x}^I}  &=   \pdv{}{\theta^{a}}\left\lbrace \sqrt{\tilde{g}_{\pi\sigma}^E \dot{x}^{\pi}\dot{x}^{\sigma}}\cdot\sqrt{\det G_0}\left[A_{\rho}^{a}-\tilde{g}^{\lambda\lambda}\tilde{g}_{\mu\rho}^EA_{\nu}^{a}\dot{x}^{\mu}\dot{x}^{\nu} \right] \right\rbrace\\ 
%%%
&=  \sqrt{\det G_0}  \Bigg\{ \pdv{\sqrt{\tilde{g}_{\pi\sigma}^E \dot{x}^{\pi}\dot{x}^{\sigma}}}{\theta^{a}}\left[A_{\rho}^{a}-\tilde{g}^{\lambda\lambda}\tilde{g}_{\mu\rho}^E A_{\nu}^{a}\dot{x}^{\mu}\dot{x}^{\nu} \right]+\\ 
& \quad +\sqrt{\tilde{g}_{\pi\sigma}^E \dot{x}^{\pi}\dot{x}^{\sigma}} \left[ \nabla^{0}_{a} A_{\rho}^{a}  -  \dot{x}^{\mu}\dot{x}^{\nu}  \left(  \pdv{\left(\tilde{g}^{\lambda\lambda}\tilde{g}_{\mu\rho}^E \right)}{\theta^{a}}  A_{\nu}^{a}+ \tilde{g}^{\lambda\lambda}\tilde{g}_{\mu\rho}^E \nabla^0_{a}A_{\nu}^{a}\right) \right] \!\! \Bigg\}
\end{split}
\end{equation}
where $\nabla^0_{a}$ is the covariant derivative w.r.t.~the round metric of the 3-sphere.

The three-dimensional geodesic $x^{\mu}$ is thus a solution if and only if this term vanishes. From eq.(\ref{eq:xiderivative}) we see that this happens if two conditions are satisfied:
\begin{enumerate}
\item $\partial_a \left( \tilde{g}_{\mu\nu}^E\right)=0$; in this way the first and third term in the second line of eq. (\ref{eq:xiderivative}) are zero. But this requirement is exactly the definition of a \textit{factorizable} metric, i.e.~a metric whose three-dimensional Einstein metric of \eqref{metricfact} does not depend on angular coordinates;
\item $\nabla_{a}^0A_{\rho}^{a}=0$; this is a gauge choice that can always be employed. It is called the \textit{de Donder} gauge. 
\end{enumerate}

Summarising our results, we have proven that  for a \textit{factorisable} six-dimensional metric in the \textit{de Donder} gauge, the full HEE defined in eq. (\ref{eq:EL6D}) reduces to:
\begin{align}
\label{eq.3D problem}
\frac{\delta \sqrt{\tilde{g}^E_{\pi\sigma}\dot{x}^{\pi}\dot{x}^{\sigma}}}{\delta x^{\rho}}-\pdv{}{\lambda}\frac{\delta \sqrt{\tilde{g}^E_{\pi\sigma}\dot{x}^{\pi}\dot{x}^{\sigma}}}{\delta \dot{x}^{\rho}} = 0 \,, 
\end{align} 
which is the same minimization problem we have to solve to find the geodesics of $\tilde{g}_{\mu\nu}^E$. Property 1 is non-trivial to be realised and might depend on a clever coordinate choice in the original six-dimensional metric. As expressed in the previous section, we will study special microstate geometries whose six-dimensional metric are indeed factorised, hence we will make extensive use of eq. \eqref{eq.3D problem} in order to find their HEE.

\section{Computing the Holographic Entanglement Entropy}\label{sec:computingHEE}

Since in sec.~\ref{sec:reducingformula} we have proven that for \textit{factorisable} geometries in the de Donder gauge the six-dimensional problem reduces  to the three-dimensional one, we can start directly with the stationary three dimensional metric
\begin{align*}
\dd s^2=  g_{tt} \dd t^2 + g_{rr}  \dd r^2+g_{yy} \dd y^2+2g_{ty} \dd t \,  \dd  y \,,
\end{align*}
where we have assumed  $\tilde{g}^E_{ry}=0=\tilde{g}^E_{rt}=0$, which is always the case for our geometries.\\
It is convenient to parametrize the geodesic $x^{\mu}(\lambda)$ in terms of \textit{proper time}, i.e.
\begin{align}
\label{eq:propertime}
\tilde{g}_{\mu\nu}^E \dot{x}^{\mu}\dot{x}^{\nu}=1 \,,
\end{align}
where $\dot{x}^{\mu} \equiv \frac{d x^{\mu}}{d \lambda}$. We recall that geodesics are defined by the minimization of the functional:
\begin{align*}
L=\int \dd \lambda \, \sqrt{g_{rr}  \dot{r}^2+g_{tt} \dot{t}^2 +g_{yy} \dot{y}^2+2g_{ty}\dot{t}\dot{y}} \equiv \int  \mathcal{L} \, \dd \lambda \Leftrightarrow 
\frac{\delta \mathcal{L}}{\delta x^{\mu}}-\pdv{}{\lambda}\frac{\delta \mathcal{L}}{\delta \dot{x}^{\mu}}=0 \,.
\end{align*}

Since we are assuming that the metric does not depend explicitly on $t$ and $y$, Euler-Lagrange equations with the respect to these coordinates gives two constants of motion, namely:
\begin{align}
\label{eq:euler-lagrange1}
\frac{\delta \mathcal{L}}{\delta \dot{t}}& =C_1 \,,\\
\label{eq:euler-lagrange2}
\frac{\delta \mathcal{L}}{\delta \dot{y}}& =C_2 \,,
\end{align}
where $C_1$ and $C_2$ are two constants of motion to be fixed later.

Solving (\ref{eq:euler-lagrange1}),(\ref{eq:euler-lagrange2}) taking into account the condition~(\ref{eq:propertime}), one finds:
\begin{align}
\label{eq:tdot}
\dot{t} &= \frac{C_1 g_{yy}-C_2 g_{ty}}{g_{tt} g_{yy}-g_{ty}^2} \,,\\
\label{eq:ydot}
\dot{y} &= \frac{C_2 g_{tt}-C_1 g_{ty}}{g_{tt} g_{yy}-g_{ty}^2} \,,\\
\label{eq:rdot}
\dot{r}^2&=\frac{1}{g_{rr}}\left[1-\left(\frac{ g_{yy} C_1^2+g_{tt} C_2^2-2 C_1 C_2 g_{ty}}{g_{tt} g_{yy}-g_{ty}^2} \right) \right] .
\end{align}
The two integration constants $C_1$ and $C_2$ are determined by the choice of boundary conditions. We are interested in a spatial region $A$ at fixed time ($t=\bar{t}$), made of an interval of length $l$.
The endpoints of the geodesic have to lie at the boundary of AdS, but for $r \to \infty$ the area diverges, as expected for the HEE in the dual CFT. So, as usual, we introduce an IR cut-off $r_0$, which we will consider as the AdS boundary for computations, and at the end we will take the results for large $r_0$. Thus boundary  conditions reads:
\begin{subequations} \label{eq:bct}
\begin{align}
0&=\int_{\bar{t}}^{\bar{t}} \dd t=\int_{\lambda_1}^{\lambda_2} \dot{t} \, \dd \lambda=2 \int_{r*}^{r_0\gg 1 } \dd r \, \frac{\dot{t}}{\dot{r}} \,,\\
\label{eq:bcy}
l&=\int_{0}^{l} \dd y=\int_{\lambda_1}^{\lambda_2} \dot{y} \,  \dd\lambda=2 \int_{r*}^{r_0\gg 1 } \dd r \,  \frac{\dot{y}}{\dot{r}}  \,,\\
\label{eq:L}
L_{\gamma}&=\int_{\lambda_1}^{\lambda_2} \dd \lambda=2 \int_{r*}^{r_0\gg 1 }  \frac{\dd r}{\dot{r}} \,,
\end{align}
\end{subequations}
where $ r_* $ is the geodesic turning point, such that $\dot{r}|_{r= r_* }=0$. In the end we compute the HEE via eq.~(\ref{enen3}). 

Up to now, we can apply our result to the generic stationary factorisable $(1,0,n)$ family of solutions; in order to give a concise and consistent explicit result, in the following we will restrict ourself to the static $(k,m,n)=(1,0,0)$ case.

\subsection{The static $k=1, n=0$ geometry as an explicit example}

We study now  the easiest, static case: the two charge $(k,m,n)=(1,0,0)$ geometry in \eqref{k=1geo}. When the metric is \textit{static} we can take a submanifold at constant $t$ and the only relevant components of the metric remain $g_{yy}(r)$ and $g_{rr}(r)$. eqs.~(\ref{eq:tdot})-(\ref{eq:rdot}) simply reduce to: 
\begin{align}
\label{eq:ydotstationary}
\frac{\dd}{\dd \lambda}\left( g_{yy}\dot{y}\right)=0\quad & \Rightarrow \quad \dot{y}=\frac{C}{g_{yy}} \,,\\
\label{eq:rdotstationary}
g_{rr}\dot{r}^2+g_{yy}\dot{y}^2=1 \quad & \Rightarrow \quad \dot{r}=\sqrt{\frac{g_{yy-C^2}}{g_{rr}g_{yy}}} \,.
\end{align}
The turning point is now defined through
\begin{align}
\label{eq:turningpointstationary}
g_{yy}( r_* )-C^2=0
\end{align}  
Boundary conditions are the same as the stationary case, apart the fact that condition (\ref{eq:bct}) is automatically satisfied and (\ref{eq:bcy}) and (\ref{eq:L}) simplify to
\begin{align}
\label{eq:lconstraintstatic}
l&=2C \int_{ r_* }^{r_0} \dd r \sqrt{\frac{g_{rr}}{g_{yy}\left(g_{yy}-C^2\right)}} \,,\\
\label{eq:Lstatic} 
L_{\gamma}&=2  \int_{ r_* }^{r_0} \dd r \sqrt{ \frac{g_{rr}g_{yy}}{g_{yy}-C^2}} \,.
\end{align} 
Let us start from determining the turning point $ r_* $, that is given by the solution of
\begin{align}
\frac{r^2}{\sqrt{Q_1 Q_5}}-C^2=0 \quad \Rightarrow \quad  r_* =\abs{C \left( Q_1 Q_5 \right)^{1/4}} \,,
\end{align}
where $r_*$ is determined by the constraint\footnote{Notice that the critical value of the entanglement shadow radius, i.e.~$r (\alpha=\pi)$, in the massless BTZ limit $a=0$ is recovered: $r^{\rm BTZ} = 2\pi^{-1}$ \cite{Balasubramanian:2014sra, Freivogel:2014lja}.}:
\begin{align}
\label{eq: omegan0}
\frac{l}{R}=\arccos\left(\frac{r_*^2-\tilde{a}^2}{r_*^2+\tilde{a}^2}\right)-\frac{r_*}{\tilde{a}}\sqrt{\frac{a^2-\tilde{a}^2}{a^2+r_*^2}} \log\left(\frac{2a^2+r_*^2-\tilde{a}^2-2\sqrt{(a^2+r_*^2)(a^2-\tilde{a}^2)}}{\tilde{a}^2+r_*^2}\right) ,
\end{align}
where
\begin{equation}
    \tilde a^2 \equiv \frac{a^4}{a_0^2} = \eta a^2 \,.
\end{equation}
For convenience we will indicate with $\alpha$ the ratio $\frac{l}{R}$, $\alpha \in [0, 2 \pi]$, and we define
\be\label{eq:falpha}
f_{a/a_0} (r)  = \alpha - \arccos\left(\frac{r^2-\tilde{a}^2}{r^2+\tilde{a}^2}\right)-\frac{r}{\tilde{a}}\sqrt{\frac{a^2-\tilde{a}^2}{a^2+r^2}} \log\left(\frac{2a^2+r^2-\tilde{a}^2-2\sqrt{(a^2+r^2)(a^2-\tilde{a}^2)}}{\tilde{a}^2+r^2}\right) .
\ee
The Entanglement Entropy of a region A, made of an interval of length $l$ is thus
\begin{align}
\label{eq:EEn0}
S_A=n_1 n_5\left\lbrace \log\left(\frac{4r_0^2}{a^2+r_*^2}\right)+\sqrt{\frac{a^2-\tilde{a}^2}{a^2+r_*^2}}\log\left(\frac{2a^2+r_*^2-\tilde{a}^2-2\sqrt{(a^2+r_*^2)(a^2-\tilde{a}^2)}}{\tilde{a}^2+r_*^2}\right)  \right\rbrace .
\end{align}

Unfortunately we were not able to invert   eq.~(\ref{eq: omegan0}) analytically and thus express $S_A$ as a function of $l$. We will discuss it in the two interesting regimes $a \gg b$ and $b \gg a$ to give an analytical intuition about the results, while we resort to a numerical computation in the generic regime. 

\subsubsection{Numerical analysis}

We resort to numerical computations to find the roots of $f_{a/a_0} (r_*)$, and we employ the \textsc{NumPy} library of \textsc{Python} for the numerical procedure.  
\begin{figure}
    \centering
    \includegraphics[scale=0.5]{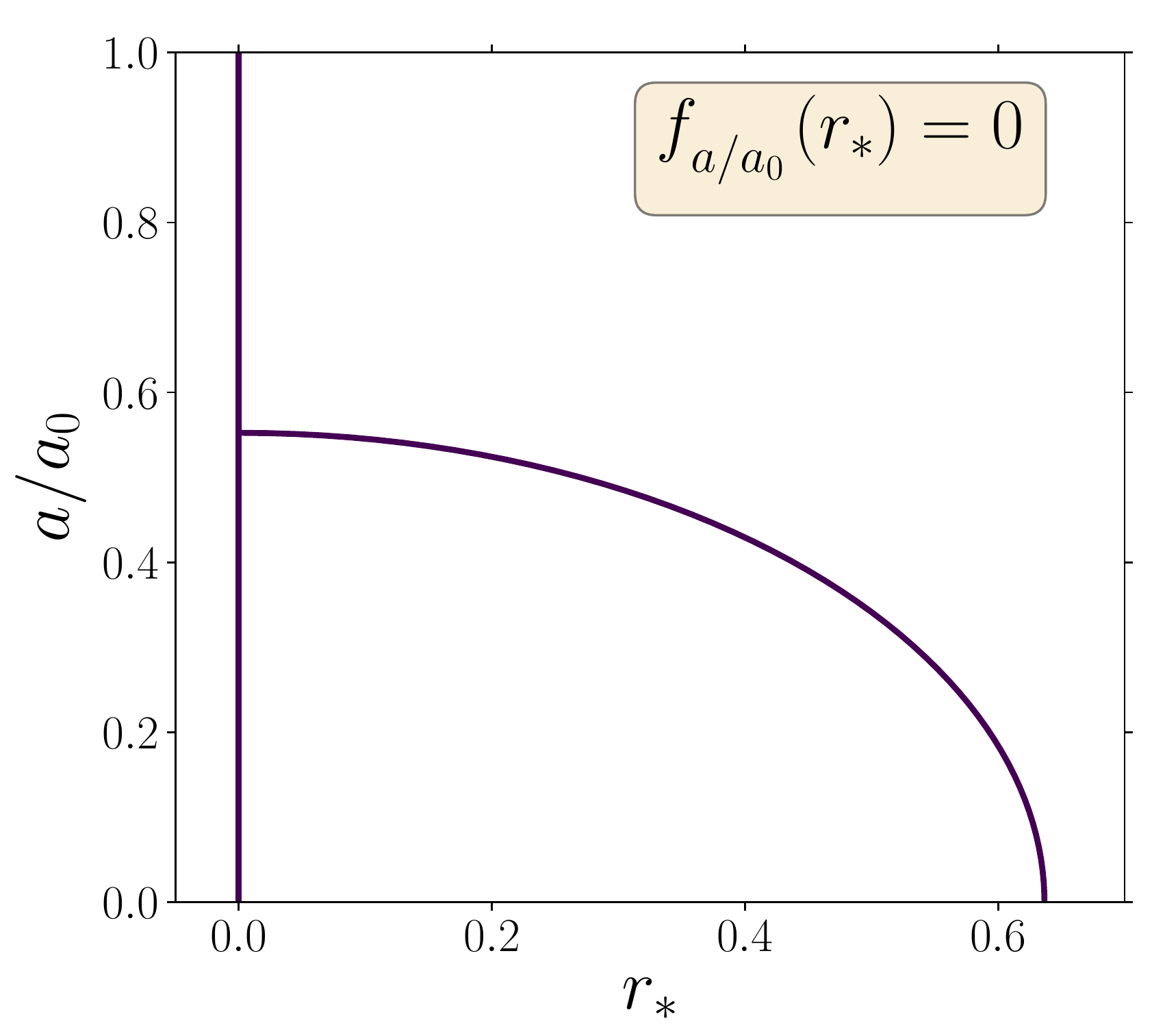}
    \caption{The plot of the roots $(r_*,a)$ of $f_{a/a_0} (r_*) $ in eq.~\eqref{eq:falpha} for $\alpha = \pi$, that is the maximal-size interval on the boundary. While $r_*=0$ is always a solution, for $a \lesssim (0.55\pm 0.01) a_0$ there exists two possible $r_*$ for any $a$. We thus need to compute the Area for both of those solutions. }
    \label{fig:shadowfig}
\end{figure}
As reported in fig.~\ref{fig:shadowfig}, while $r_*=0$ is a solution for all values of $a$, there exists a range of values for $a$ where the second branch of possible roots exists. This range is approximately $a\in (0, 0.55)a_0$ for the maximal-size boundary interval $\alpha = \pi$ and on that branch $r_*>0$, signaling a possible existence of an entanglement shadow on these geometries. This means that, for $a \lesssim 0.55 \, a_0$, there are two extremal surfaces that are attached to the boundary. In order to understand which one of these two extremal surfaces is minimal, we need to (numerically) compute the entanglement entropy. \begin{figure}
    \centering
    \includegraphics[scale=0.55]{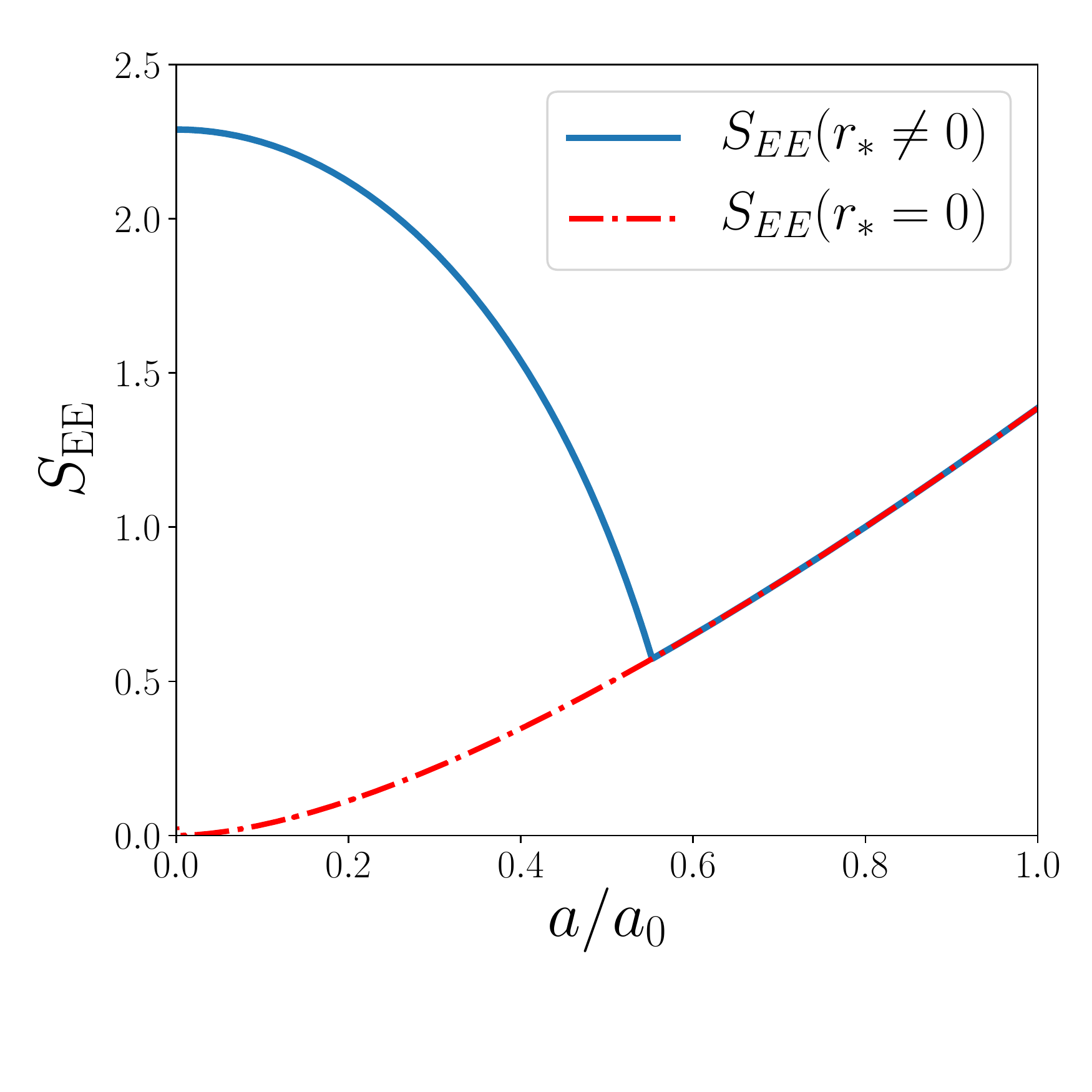}
    \caption{The entanglement entropy for the maximal-size $\alpha=\pi$ boundary interval. We reported in blue the Area of the non-trivial branch of roots for $f_{a/a_0} (r_*)$, while the red dash-dotted curve is the Area of the curve with $r_* =0$. }
    \label{fig:SEE}
\end{figure}
Surprisingly, the minimal surface is always the curve with vanishing entanglement shadow radius, for which we have
\begin{equation}
    S_{\rm EE} = n_1 n_5\left\lbrace \log\left(\frac{4r_0^2}{a^2}\right)+\sqrt{1-\frac{a^2}{a_0^2}} \, \log\left(\frac{2a_0^2 -a^2 -2a_0 \sqrt{ a_0^2 - a^2 }}{a^2}\right)  \right\rbrace , 
\end{equation}
implying that, on pure microstate geometries, \emph{there is no entanglement shadow} whatsoever. This is somehow expected, since these geometries are dual to pure CFT states, and  it seems unnatural to find an obstruction to bulk reconstruction on the supergravity side for these states. Since the HEE seems to be well defined up to $a=0$, one may wonder how it is possible to recover the standard massless BTZ result, that has a non-vanishing shadow. As explained above, this is due to the fact that, in the strict $a=0$ case, for the na\"ive massless BTZ geometry, the $r=0$ point of the space-time is the location of the (string-sized) black hole horizon; this means that the geodesic that passes through $r=0$ has no physical meaning, and has to be discarded\footnote{Actually, setting $a=0$ from the beginning, the result $r_*=0$ doesn't even exist. This is because the two limits, i.e.~$a\to 0$ and $r\to 0$, do not commute.}. This means that only the HEE for the curve with non-vanishing $r_*$ is physically viable and thus we reproduce the results of \cite{Freivogel:2014lja}. 

We will now move to compute analytically the entanglement entropy in the asymptotic regions $a \gg b$ and $a\ll b$, to have at least some approximate analytical expressions for the entanglement entropy in these two regions, that represent perturbative corrections to the vacuum AdS$_3$ and BTZ$_3$ case, respectively.

\subsubsection{The $b \ll a$ limit}
In the $a/a_0 \to 1$ or $b \ll a$ case,  the problem is exactly the same as \textit{pure} AdS$_3$\footnote{In this approximation up to $O(b^2)$ the metric is:
$$
\dd s^2 \simeq \sqrt{Q_1 Q_5} \frac{1}{r^2+a_0^2} \, \dd r^2-\frac{r^2+a_0^2-b^2}{\sqrt{Q_1 Q_5}} \, \dd t^2+\frac{r^2}{\sqrt{Q_1 Q_5}} \, \dd y^2 \,.
$$
there is no difference between the computation in this metric and that one for \textit{pure AdS} since at order $b^2$ the only term which receive corrections is $g_{tt}$, which never enters the  calculations.} up to the order $b^2$, so we have
\begin{align}\label{eq_shadowk1zero}
r_* (\alpha) = a_0 \cot \frac{\alpha}{2} +\mathcal{O}(b^4) ,
\end{align}
and then the HEE is the same as the pure AdS$_3$ case 
\begin{align}
S_A=2 n_1 n_5  \log \left[\frac{2 r_0}{a_0} \sin \left(\frac{\alpha }{2}\right)\right]+\mathcal{O}(b^4) .
\end{align}
This is obvious since we have seen that, up to a certain value of $a$, $r_*=0$ is the only solution to the equation $f_{a/a_0} (r_*) =0$.

\subsubsection{The $b \gg a$ limit}

In the $a/a_0 \ll 1$ or $b \gg a$ regime the geometry approaches the na\"ive massless BTZ geometry. The eq. \eqref{eq: omegan0} for $r_*>0$ simplifies a lot and allows us to invert the relation 
\be
r_*  (\alpha) = a_0 \left(\frac{2}{\alpha}-\frac{a^2}{a_0^2}\frac{\alpha}{3} \right)+\mathcal{O}\left(\frac{a^4}{a_0^4}\right) ,
\ee
so that the entanglement entropy becomes 
\begin{align}
S_A \simeq n_1 n_5 \left[ \log \left(\frac{\alpha ^2 r_0^2}{a_0^2}\right)-\frac{a^2}{a_0^2}\frac{\alpha ^2}{6}\right] ,
\end{align}
where we recall that $r_0$ is the IR cut-off. We have then reproduced the usual BTZ result plus corrections that reduces the size of the entangling shadows. This result is the approximate expansion near $a=0$ of the blue curve of fig.~\ref{fig:SEE}, that turns out not to be the minimal surface. It is still worth to compute the correction to it, since one may wonder what happens in geometries with small corrections w.r.t.~the na\"ive black hole geometry.

\section{The Holographic Complexity on microstate geometries} \label{sec:complexity=volume}

In the last years, one new observable gained attention in the black hole community, as well as in the quantum information one: the (holographic) complexity \cite{Susskind:2014moa, Susskind:2014rva, Alishahiha:2015rta, Brown:2015bva, Brown:2015lvg, Chapman:2016hwi, Carmi:2016wjl, Reynolds:2016rvl, Carmi:2017jqz, Reynolds:2017jfs, Chapman:2018dem, Susskind:2018vql, Susskind:2018pmk, Chapman:2018hou, Balasubramanian:2018hsu, Goto:2018iay, Bernamonti:2019zyy, Ross:2019rtu}. The two proposals for computing holographically the complexity are the ``complexity = action" or ``C=A'' proposal, where the complexity is proposed to be dual to the on-shell action on the Wheeler-de Witt patch, while the second one is the ``complexity = volume'' or ``C=V'' proposal, where the complexity is proposed to be dual to the volume of the maximal spacelike co-dimension 1 surface in the bulk. 

We will study the latter for its simplicity in the case at hand. The first thing to notice  is that, for factorisable metric in the de Donder Gauge, one can easily generalise the computation of sec.~\ref{sec:reducingformula} and prove that one may reduce to compute the volume of the maximal spacelike slice in the Euclidean three dimensional metric. 

\subsection{The Complexity = Volume of the two-charge geometry}
If we assume that the (holographic) Complexity of the state on a boundary time-slice $\Omega$ is given by the volume of the extremal time-slice ${\cal B}$ (such that $\Omega = \pd {\cal B}$) via the formula
\begin{equation}
    {\cal C}_V [\Omega] = \max_{\Omega = \pd {\cal B}} \left[ \frac{{\cal V} [{\cal B}]}{G_N \ell}  \right],
\end{equation}
we have that, in the $(1,0,0)$ geometry, the induced 2-dimensional geometry is 
\be
\dd s_{\cal B}^2 = \frac{r^2+ a_0^2 \eta^2}{(r^2+ a_0^2 \eta)^2} \, \dd r^2 + r^2 \dd \sigma \,, 
\ee
where we recall that $\eta = a^2/a_0^2$, so that it is straightforward to compute the holographic complexity as  
\be
{\cal C}_V (\eta) = \int_0^{r_0} \dd r  \int_0^{2\pi} \dd \sigma \, \sqrt{h} = 2\pi r_0 - 2 \pi a_0 \sqrt{\eta}  \left(\sqrt{\eta} + \sqrt{1-\eta} \, \arccos \sqrt{\eta}  \right) ,
\ee
where we set $G_3 =1$ and we recall that $\ell=a_0$.
As usual when computing holographic quantities in AdS, the value of the complexity turns out to be divergent; this is why we introduced a regularising scale $r_0$. We can then compute the regularised Complexity (or ``complexity of formation'' \cite{Chapman:2016hwi}) by subtracting the ``vacuum'' contribution, i.e.~${\cal C}_V^{\rm vac} \equiv {\cal C}_V (\eta=1)$\footnote{Notice that, for $\eta=0$, if we use the use the same notation of \cite{Chapman:2016hwi}, we reproduce their result eq.~(5.17) for massless BTZ.}:
\be
{\cal C}_V^{\rm reg} (\eta)  \equiv {\cal C}_V (\eta)  - {\cal C}_V^{\rm vac}  =   2\pi a_0 \left(1-\eta- \sqrt{\eta(1-\eta)} \,  \arccos \sqrt{\eta} \right) .
\ee
This has the correct behaviour; when $\eta \to 1$, i.e.~when we are approaching  the most ``simple'' state, ${\cal C}_V^{\rm reg} (\eta) \to 0$; also, when $\eta \to 0$, i.e.~when we are approaching the most ``complex'' state, ${\cal C}_V^{\rm reg} (\eta) \to 2\pi a_0$. In other words:
\begin{itemize}
    \item when $\eta \to 1$ so that the $\frac{1}{4}$-BPS state \eqref{eq:2chargeCFTstate} approaches $\psi_{\{N^{(++)}, N^{(00)}\}} \sim \left|++\right\rangle_1^N$, we have that ${\cal C}_V^{\rm reg} (\eta) \to 0$; 
    \item when $\eta \to 0$ so that the $\frac{1}{4}$-BPS state approaches $\psi_{\{N^{(++)}, N^{(00)}\}} \sim |00\rangle_1^N$, we have that ${\cal C}_V^{\rm reg} (\eta) \to 2\pi a_0$;
\end{itemize}
we can also see from fig.~\ref{fig:Cvreg} that the behaviour of ${\cal C}_V^{\rm reg} (\eta) $ is indeed monotonic.
\begin{figure}
    \centering
    \includegraphics[scale=0.55]{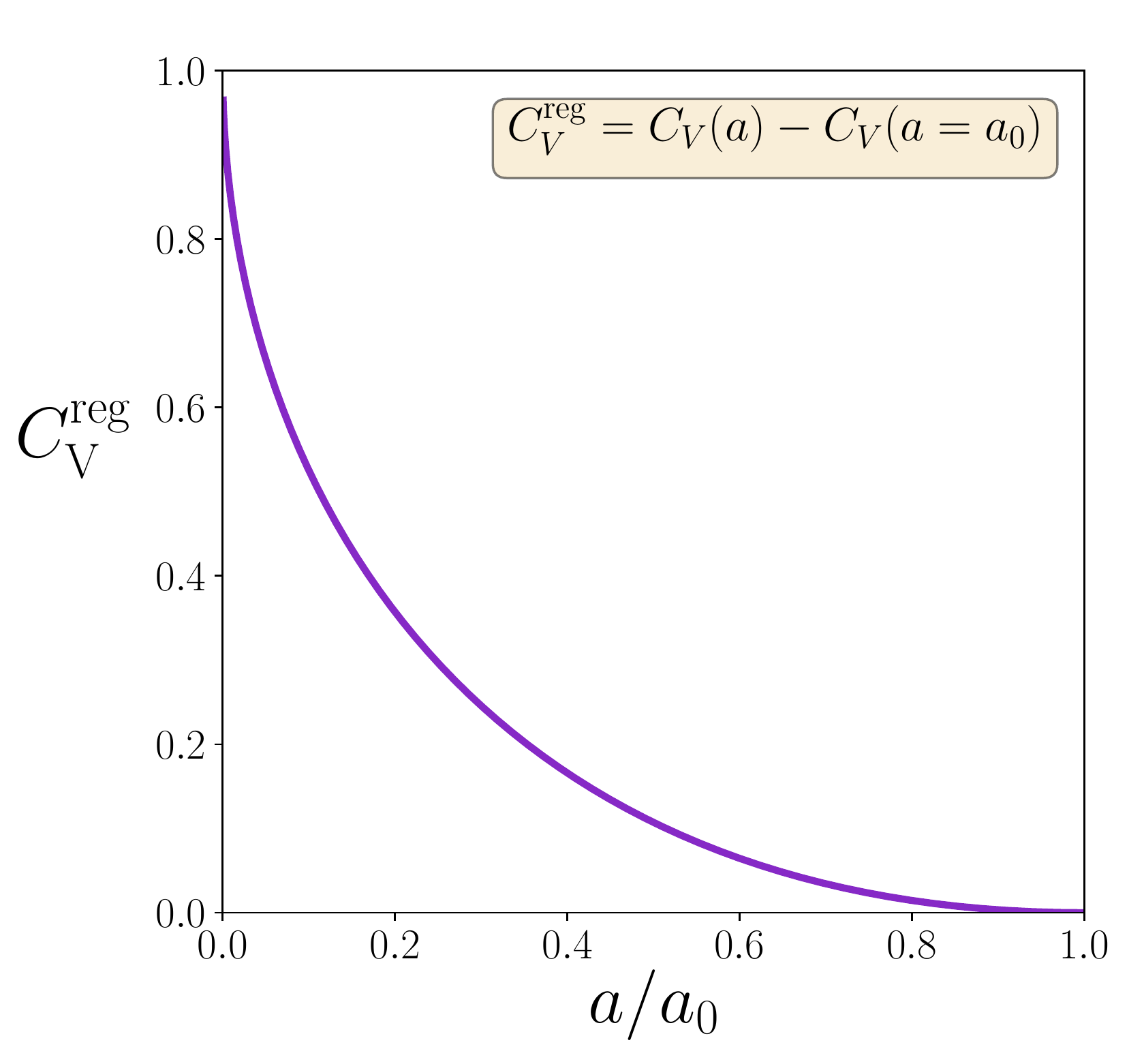}
    \caption{The holographic complexity of formation of the $\frac{1}{4}$-BPS state \eqref{eq:2chargeCFTstate}, expressed in units of $2\pi a_0$. }
    \label{fig:Cvreg}
\end{figure}
%One may wonder why, in moving on the $a/a_0$ line, we have a finite complexity of formation; if the putative CFT dual of the complexity is the \emph{Nielsen's Circuit Complexity}, that is the geodesic length in circuit space that is related to the minimum number of (quantum logical) gates to synthesise a desired unitary operator \cite{Nielsen1133, Nielsen:PhysRevA73, Jefferson:2017sdb, Guo:2018kzl, Chapman:2017rqy, Camargo:2019isp, Bhattacharyya:2019kvj}, one may wonder why we have a finite geodesic length that connects a pure state to a thermal state; again, we need to remark that, even if \emph{in the supergravity limit} the $a=0$ is indistinguishable to the na\"ive massless BTZ geometry, at the full-fledged string-theoretical level the two geometry \emph{are different}, with the former that is dual to the pure state $|00\rangle_1^N$ that still has no singularity nor horizon. It is thus indeed expected that the ``distance'' between the state $\left|++\right\rangle_1^N$ and the state $|00\rangle_1^N$ to be finite. 

\section{Discussion}\label{sec:discussion}

In this paper, we have discussed the holographic entanglement entropy for a one-parameter family of microstate geometries that are asymptotically AdS$_3 \times \mathbb{S}^3 \times \mathbb{T}^4$. This family of geometries is controlled by the dimensionless parameter $\eta\in (0,1)$ and has the peculiarity that, for $\eta \to 0$, it approaches the massless BTZ geometry, while for $\eta \to 1$ it approaches the vacuum global AdS$_3$ geometry.

We have first shown that, for factorisable six-dimensional geometries in the de Donder gauge, the problem of computing the HEE reduces to a pure three-dimensional one; after that, we have resort to a numerical procedure to compute the minimal-area geodesic, finding that for all solutions of the one-parameter family we have considered there is no entanglement shadow. The key point is that the non-linear equation, relating the size of the boundary interval $l = \alpha R$ to the turning point radius $r_*$ (and thus to the entanglement shadow), admits two solutions in the $\eta\in (0, 0.55)$ region: one is $r_* =0$ while the other has $r_*>0$. When computing the length of the associated geodesics, it turns out that the latter is always bigger than the former, thus indicating that this family of geometries never shows any entanglement shadows. 

It seems to be in contrast with the result for the massless BTZ case, which does show entanglement shadow \cite{Freivogel:2014lja}; what happens is that the result is recovered by a somewhat non-trivial mechanism. In fact, for $a=0$, the geodesic with $r_*=0$ does not exist since the $r=0$ is the location of the (degenerate) black hole horizon; this means that in this case there exists only one geodesic, namely the one with $r_*>0$, and we thus reproduce the known result for the massless BTZ black hole. This suggests that the na\"ive BTZ geometry (dual to a thermal state) is strikingly different from the microstate one (dual to the pure state $|00\rangle_1^N$), even if the difference is not captured by the supergravity regime. 

We have also computed the holographic complexity via the ``complexity = volume'' duality, since the results of sec.~\ref{sec:reducingformula} can be easily generalised to the computations of the volume of the extremal slice for this family, finding a monotonic function in the parameter $\eta$, that interpolates smoothly the known results for vacuum global AdS$_3$ with the one for massless BTZ case \cite{Chapman:2016hwi}.

One may wonder to extend the results of this paper among many directions; one may try to use the results of sec.~\ref{sec:reducingformula} to stationary but not static geometries, as the $\frac{1}{8}$-BPS factorisable geometries of the $(1,0,n)$ superstrata, where one needs to refine the numerical procedure, as well as trying to compute the complexity via the ``complexity = action'', trying to understand if it is possible to reduce the problem from a six-dimensional to a three-dimensional one even in this case. In the case of non-stationary geometries that are separable, e.g.~the $(1,0,n)$ superstrata, the holographic entanglement entropy computations become more involved, and different numerical procedure may be needed, but the basic ingredients that are necessary to achieve that are the same employed in this paper, and we expect similar results to hold, i.e.~that any entanglement shadow emerges, even for finite $n$.

\vspace{7mm}
 \noindent {\large \textbf{Acknowledgements} }
 
 The authors want to thank Erik Tonni, Iosif Bena, Monica Guica, Ruben Monten, Andrea Galliani and Davide Billo for discussions. 
 The authors want to especially thank Rodolfo Russo and Stefano Giusto for their helpful insights, advises and comments on the draft. AB is deeply indebted with Jacopo Sisti for stimulating and illuminating discussions and for his comments. The authors want to thank Federico Pobbe for his crucial help with \textsc{Python} language. AB is supported by the Swedish Research Council grant number 2015-05333. GF is supported  by the Knut and Alice Wallenberg Foundation under grant KAW 2016.0129 and by VR grant 2018-04438.
\vspace{5mm}

\appendix

\section{Holographic Entanglement Entropy and Complexity for the BTZ black hole}
In this appendix we briefly review the computations of Holographic Entanglement Entropy (and the emergence of a non-vanishing shadow) and the complexity for a massive BTZ black hole, taking in the end the limit $M\to 0$ limit to look at the massless BTZ results. The metric for the (non-rotating) massive BTZ black hole is
\begin{equation}
    \dd s^2 = - \frac{r^2 - r_h^2}{\ell^2} \, \dd t^2 + \frac{\ell^2}{r^2 - r_h^2} \, \dd r^2 + r^2 \dd y^2 \,,
\end{equation}
where we have explicitly inserted the AdS$_3$ radius $\ell$, and where $y \sim y + 2\pi$.

\subsection{The Holographic Entanglement Entropy and Shadow}
Following the lines of sec.~\ref{sec:computingHEE}, we have that $g_{yy} (r_*)- C^2 =0$ defines $C= r_*$, so that, for an interval of size $L/R = \alpha$
\begin{equation}
    \begin{split}
        \alpha &=2 C \int_{r_*}^{r_0} \dd r \, \sqrt{\frac{g_{rr}}{g_{yy} (g_{yy}-C^2)}} \\
        &= \frac{2\ell}{r_h} {\rm arctanh} \left[ \frac{r_h}{r_*}\right] , \\
        %%%
        {\cal A} &= 2 \int_{r_*}^{r_0} \dd r \,\sqrt{\frac{g_{rr} g_{yy}}{g_{yy}-C^2}} \\
        &=\ell \log \frac{r^2}{\ell^2} + \ell \log \left[ \frac{4 \ell^2}{r_*^2-r_h^2}\right] .
    \end{split}
\end{equation}
The first line can be inverted to find $r_* =r_*(\alpha)$ as
\begin{equation}
    r_* (\alpha) = r_h \coth \left[ \frac{r_h}{\ell} \, \frac{\alpha}{2} \right] , 
\end{equation}
that immediately tells us that, taking the largest possible interval on the boundary $\alpha=\pi$, we have a non-vanishing shadow 
\begin{equation}\label{eq:BTZMshadow}
    r_{\rm shadow} =  r_h \coth \left[ \frac{r_h}{\ell} \, \frac{\pi}{2} \right] ,
\end{equation}
matching the results of \cite{Freivogel:2014lja}; also, in the massless limit $r_h\to 0$, we have
\begin{equation}\label{eq:BTZM=0shadow}
    r_{\rm shadow}^{\rm M=0} = \frac{2}{\pi} \, \ell \,.
\end{equation}
Now, plugging \eqref{eq:BTZMshadow} into the formula for the Area, we get 
\begin{equation}
    S_{\rm HEE} = \frac{\cal A}{4 G_3} = \frac{\ell}{4 G} \left\{ \log \frac{r^2}{\ell^2} + \log \left[ \frac{4 \ell^2}{r_h^2} \sinh  \left( \frac{r_h}{\ell} \, \frac{\alpha}{2} \right) \right] \right\} ,
\end{equation}
matching the results of \cite{Cadoni:2009tk}. Again, in the massless limit 
\begin{equation}
    S_{\rm HEE}^{M=0} = \frac{\cal A}{4 G_3} = \frac{\ell}{4 G} \left\{ \log \frac{r^2}{\ell^2} + \log \alpha^2 \right\} .
\end{equation}

\subsection{The Complexity = Volume}
Since the metric is static, the maximal spacelike slice is obtained by setting $t=$const.~and then 
\begin{equation}
\begin{split}
        {\cal V}_{\rm BTZ} &= 2 \pi \int_{r_h}^{r_0} \dd r \, \sqrt{h} = \int_{r_h}^{r_0} \dd r \, \frac{\ell r}{\sqrt{r^2 - r_h^2}} \\
        &= 2 \pi \ell r_0 \,.
\end{split}
\end{equation}
The AdS vacuum term is obtained by a simple trick, i.e.~by Wick rotating $r_h \to i \ell$ and integrating from 0 to the cut-off scale $r_0$, so that it gives
\begin{equation}
\begin{split}
        {\cal V}_{\rm vac} &= 2 \pi \int_{0}^{r_0} \dd r \, \sqrt{h_{\rm vac}} = \int_{0}^{r_0} \dd r \, \frac{\ell r}{\sqrt{r^2 + \ell^2}} \\
        &= 2 \pi \ell (r- \ell)\,.
\end{split}
\end{equation}
We can now easily compute the complexity of formation, that is 
\begin{equation}
    {\cal C}_V^{\rm reg} = \frac{1}{G_3 \ell} \left[  {\cal V}_{\rm  BTZ}-{\cal V}_{\rm vac} \right] = \frac{2 \pi \ell}{G_3} \,,
\end{equation}
that is a constant independent of the mass of the BTZ black hole \cite{Chapman:2016hwi}. 

\newpage

\providecommand{\href}[2]{#2}\begingroup\raggedright\endgroup
 

\begin{thebibliography}{10}

\bibitem{Calabrese:2004eu}
P.~Calabrese and J.~L. Cardy, ``{Entanglement entropy and quantum field
  theory},'' {\em J. Stat. Mech.} {\bf 0406} (2004) P06002,
\href{http://arXiv.org/abs/hep-th/0405152}{{\tt hep-th/0405152}}.
%%CITATION = HEP-TH/0405152;%%.

\bibitem{Susskind:2014rva}
L.~Susskind, ``{Computational Complexity and Black Hole Horizons},'' {\em
  Fortsch. Phys.} {\bf 64} (2016) 44--48,
  \href{http://arXiv.org/abs/1403.5695}{{\tt 1403.5695}}.
[Fortsch. Phys.64,24(2016)].
%%CITATION = ARXIV:1403.5695;%%.

\bibitem{Susskind:2014moa}
L.~Susskind, ``{Entanglement is not enough},'' {\em Fortsch. Phys.} {\bf 64}
  (2016) 49--71,
\href{http://arXiv.org/abs/1411.0690}{{\tt 1411.0690}}.
%%CITATION = ARXIV:1411.0690;%%.

\bibitem{Stanford:2014jda}
D.~Stanford and L.~Susskind, ``{Complexity and Shock Wave Geometries},'' {\em
  Phys. Rev.} {\bf D90} (2014), no.~12, 126007,
\href{http://arXiv.org/abs/1406.2678}{{\tt 1406.2678}}.
%%CITATION = ARXIV:1406.2678;%%.

\bibitem{Ryu:2006bv}
S.~Ryu and T.~Takayanagi, ``{Holographic derivation of entanglement entropy
  from AdS/CFT},'' {\em Phys. Rev. Lett.} {\bf 96} (2006) 181602,
\href{http://arXiv.org/abs/hep-th/0603001}{{\tt hep-th/0603001}}.
%%CITATION = HEP-TH/0603001;%%.

\bibitem{Ryu:2006ef}
S.~Ryu and T.~Takayanagi, ``{Aspects of Holographic Entanglement Entropy},''
  {\em JHEP} {\bf 08} (2006) 045,
\href{http://arXiv.org/abs/hep-th/0605073}{{\tt hep-th/0605073}}.
%%CITATION = HEP-TH/0605073;%%.

\bibitem{Hubeny:2007xt}
V.~E. Hubeny, M.~Rangamani, and T.~Takayanagi, ``{A Covariant holographic
  entanglement entropy proposal},'' {\em JHEP} {\bf 07} (2007) 062,
\href{http://arXiv.org/abs/0705.0016}{{\tt 0705.0016}}.
%%CITATION = ARXIV:0705.0016;%%.

\bibitem{Alishahiha:2015rta}
M.~Alishahiha, ``{Holographic Complexity},'' {\em Phys. Rev.} {\bf D92} (2015),
  no.~12, 126009,
\href{http://arXiv.org/abs/1509.06614}{{\tt 1509.06614}}.
%%CITATION = ARXIV:1509.06614;%%.

\bibitem{Brown:2015bva}
A.~R. Brown, D.~A. Roberts, L.~Susskind, B.~Swingle, and Y.~Zhao,
  ``{Holographic Complexity Equals Bulk Action?},'' {\em Phys. Rev. Lett.} {\bf
  116} (2016), no.~19, 191301,
\href{http://arXiv.org/abs/1509.07876}{{\tt 1509.07876}}.
%%CITATION = ARXIV:1509.07876;%%.

\bibitem{Brown:2015lvg}
A.~R. Brown, D.~A. Roberts, L.~Susskind, B.~Swingle, and Y.~Zhao,
  ``{Complexity, action, and black holes},'' {\em Phys. Rev.} {\bf D93} (2016),
  no.~8, 086006,
\href{http://arXiv.org/abs/1512.04993}{{\tt 1512.04993}}.
%%CITATION = ARXIV:1512.04993;%%.

\bibitem{Lehner:2016vdi}
L.~Lehner, R.~C. Myers, E.~Poisson, and R.~D. Sorkin, ``{Gravitational action
  with null boundaries},'' {\em Phys. Rev.} {\bf D94} (2016), no.~8, 084046,
\href{http://arXiv.org/abs/1609.00207}{{\tt 1609.00207}}.
%%CITATION = ARXIV:1609.00207;%%.

\bibitem{Chapman:2016hwi}
S.~Chapman, H.~Marrochio, and R.~C. Myers, ``{Complexity of Formation in
  Holography},'' {\em JHEP} {\bf 01} (2017) 062,
\href{http://arXiv.org/abs/1610.08063}{{\tt 1610.08063}}.
%%CITATION = ARXIV:1610.08063;%%.

\bibitem{Calabrese:2009qy}
P.~Calabrese and J.~Cardy, ``{Entanglement entropy and conformal field
  theory},'' {\em J. Phys.} {\bf A42} (2009) 504005,
\href{http://arXiv.org/abs/0905.4013}{{\tt 0905.4013}}.
%%CITATION = ARXIV:0905.4013;%%.

\bibitem{Rangamani:2016dms}
M.~Rangamani and T.~Takayanagi, ``{Holographic Entanglement Entropy},'' {\em
  Lect. Notes Phys.} {\bf 931} (2017) pp.1--246,
\href{http://arXiv.org/abs/1609.01287}{{\tt 1609.01287}}.
%%CITATION = ARXIV:1609.01287;%%.

\bibitem{Michel:2018yta}
B.~Michel and A.~Puhm, ``{Corrections in the relative entropy of black hole
  microstates},'' {\em JHEP} {\bf 07} (2018) 179,
\href{http://arXiv.org/abs/1801.02615}{{\tt 1801.02615}}.
%%CITATION = ARXIV:1801.02615;%%.

\bibitem{Seiberg:1999xz}
N.~Seiberg and E.~Witten, ``{The D1 / D5 system and singular CFT},'' {\em JHEP}
  {\bf 04} (1999) 017,
\href{http://arXiv.org/abs/hep-th/9903224}{{\tt hep-th/9903224}}.
%%CITATION = HEP-TH/9903224;%%.

\bibitem{David:2002wn}
J.~R. David, G.~Mandal, and S.~R. Wadia, ``{Microscopic formulation of black
  holes in string theory},'' {\em Phys. Rept.} {\bf 369} (2002) 549--686,
\href{http://arXiv.org/abs/hep-th/0203048}{{\tt hep-th/0203048}}.
%%CITATION = HEP-TH/0203048;%%.

\bibitem{Avery:2010qw}
S.~G. Avery, {\em {Using the D1D5 CFT to Understand Black Holes}}.
\newblock PhD thesis, Ohio State U., 2010.
\newblock
\href{http://arXiv.org/abs/1012.0072}{{\tt 1012.0072}}.
\newblock
%%CITATION = ARXIV:1012.0072;%%.

\bibitem{Balasubramanian:2005qu}
V.~Balasubramanian, P.~Kraus, and M.~Shigemori, ``{Massless black holes and
  black rings as effective geometries of the D1-D5 system},'' {\em Class.
  Quant. Grav.} {\bf 22} (2005) 4803--4838,
\href{http://arXiv.org/abs/hep-th/0508110}{{\tt hep-th/0508110}}.
%%CITATION = HEP-TH/0508110;%%.

\bibitem{Bombini:2017sge}
A.~Bombini, A.~Galliani, S.~Giusto, E.~Moscato, and R.~Russo, ``{Unitary
  4-point correlators from classical geometries},'' {\em Eur. Phys. J.} {\bf
  C78} (2018), no.~1, 8,
\href{http://arXiv.org/abs/1710.06820}{{\tt 1710.06820}}.
%%CITATION = ARXIV:1710.06820;%%.

\bibitem{Hubeny:2013gta}
V.~E. Hubeny, H.~Maxfield, M.~Rangamani, and E.~Tonni, ``{Holographic
  entanglement plateaux},'' {\em JHEP} {\bf 08} (2013) 092,
\href{http://arXiv.org/abs/1306.4004}{{\tt 1306.4004}}.
%%CITATION = ARXIV:1306.4004;%%.

\bibitem{Freivogel:2014lja}
B.~Freivogel, R.~Jefferson, L.~Kabir, B.~Mosk, and I.-S. Yang, ``{Casting
  Shadows on Holographic Reconstruction},'' {\em Phys. Rev.} {\bf D91} (2015),
  no.~8, 086013,
\href{http://arXiv.org/abs/1412.5175}{{\tt 1412.5175}}.
%%CITATION = ARXIV:1412.5175;%%.

\bibitem{Balasubramanian:2014sra}
V.~Balasubramanian, B.~D. Chowdhury, B.~Czech, and J.~de~Boer, ``{Entwinement
  and the emergence of spacetime},'' {\em JHEP} {\bf 01} (2015) 048,
\href{http://arXiv.org/abs/1406.5859}{{\tt 1406.5859}}.
%%CITATION = ARXIV:1406.5859;%%.

\bibitem{Lunin:2001jy}
O.~Lunin and S.~D. Mathur, ``{AdS/CFT duality and the black hole information
  paradox},'' {\em Nucl. Phys.} {\bf B623} (2002) 342--394,
\href{http://arXiv.org/abs/hep-th/0109154}{{\tt hep-th/0109154}}.
%%CITATION = HEP-TH/0109154;%%.

\bibitem{Lunin:2002iz}
O.~Lunin, J.~M. Maldacena, and L.~Maoz, ``{Gravity solutions for the D1-D5
  system with angular momentum},''
\href{http://arXiv.org/abs/hep-th/0212210}{{\tt hep-th/0212210}}.
%%CITATION = HEP-TH/0212210;%%.

\bibitem{Mathur:2003hj}
S.~D. Mathur, A.~Saxena, and Y.~K. Srivastava, ``{Constructing `hair' for the
  three charge hole},'' {\em Nucl. Phys.} {\bf B680} (2004) 415--449,
\href{http://arXiv.org/abs/hep-th/0311092}{{\tt hep-th/0311092}}.
%%CITATION = HEP-TH/0311092;%%.

\bibitem{Lunin:2004uu}
O.~Lunin, ``{Adding momentum to D1-D5 system},'' {\em JHEP} {\bf 04} (2004)
  054,
\href{http://arXiv.org/abs/hep-th/0404006}{{\tt hep-th/0404006}}.
%%CITATION = HEP-TH/0404006;%%.

\bibitem{Giusto:2004ip}
S.~Giusto, S.~D. Mathur, and A.~Saxena, ``{3-charge geometries and their CFT
  duals},'' {\em Nucl. Phys.} {\bf B710} (2005) 425--463,
\href{http://arXiv.org/abs/hep-th/0406103}{{\tt hep-th/0406103}}.
%%CITATION = HEP-TH/0406103;%%.

\bibitem{Giusto:2004id}
S.~Giusto, S.~D. Mathur, and A.~Saxena, ``{Dual geometries for a set of
  3-charge microstates},'' {\em Nucl. Phys.} {\bf B701} (2004) 357--379,
\href{http://arXiv.org/abs/hep-th/0405017}{{\tt hep-th/0405017}}.
%%CITATION = HEP-TH/0405017;%%.

\bibitem{Skenderis:2006ah}
K.~Skenderis and M.~Taylor, ``{Fuzzball solutions and D1-D5 microstates},''
  {\em Phys. Rev. Lett.} {\bf 98} (2007) 071601,
\href{http://arXiv.org/abs/hep-th/0609154}{{\tt hep-th/0609154}}.
%%CITATION = HEP-TH/0609154;%%.

\bibitem{Kanitscheider:2006zf}
I.~Kanitscheider, K.~Skenderis, and M.~Taylor, ``{Holographic anatomy of
  fuzzballs},'' {\em JHEP} {\bf 04} (2007) 023,
\href{http://arXiv.org/abs/hep-th/0611171}{{\tt hep-th/0611171}}.
%%CITATION = HEP-TH/0611171;%%.

\bibitem{Skenderis:2007yb}
K.~Skenderis and M.~Taylor, ``{Anatomy of bubbling solutions},'' {\em JHEP}
  {\bf 09} (2007) 019,
\href{http://arXiv.org/abs/0706.0216}{{\tt 0706.0216}}.
%%CITATION = ARXIV:0706.0216;%%.

\bibitem{Kanitscheider:2007wq}
I.~Kanitscheider, K.~Skenderis, and M.~Taylor, ``{Fuzzballs with internal
  excitations},'' {\em JHEP} {\bf 06} (2007) 056,
\href{http://arXiv.org/abs/0704.0690}{{\tt 0704.0690}}.
%%CITATION = ARXIV:0704.0690;%%.

\bibitem{Skenderis:2008qn}
K.~Skenderis and M.~Taylor, ``{The fuzzball proposal for black holes},'' {\em
  Phys. Rept.} {\bf 467} (2008) 117--171,
\href{http://arXiv.org/abs/0804.0552}{{\tt 0804.0552}}.
%%CITATION = ARXIV:0804.0552;%%.

\bibitem{Mathur:2011gz}
S.~D. Mathur and D.~Turton, ``{Microstates at the boundary of AdS},'' {\em
  JHEP} {\bf 1205} (2012) 014,
\href{http://arXiv.org/abs/1112.6413}{{\tt 1112.6413}}.
%%CITATION = ARXIV:1112.6413;%%.

\bibitem{Mathur:2012tj}
S.~D. Mathur and D.~Turton, ``{Momentum-carrying waves on D1-D5 microstate
  geometries},'' {\em Nucl.Phys.} {\bf B862} (2012) 764--780,
\href{http://arXiv.org/abs/1202.6421}{{\tt 1202.6421}}.
%%CITATION = ARXIV:1202.6421;%%.

\bibitem{Lunin:2012gp}
O.~Lunin, S.~D. Mathur, and D.~Turton, ``{Adding momentum to supersymmetric
  geometries},'' {\em Nucl.Phys.} {\bf B868} (2013) 383--415,
\href{http://arXiv.org/abs/1208.1770}{{\tt 1208.1770}}.
%%CITATION = ARXIV:1208.1770;%%.

\bibitem{Giusto:2013rxa}
S.~Giusto, L.~Martucci, M.~Petrini, and R.~Russo, ``{6D microstate geometries
  from 10D structures},'' {\em Nucl. Phys.} {\bf B876} (2013) 509--555,
\href{http://arXiv.org/abs/1306.1745}{{\tt 1306.1745}}.
%%CITATION = ARXIV:1306.1745;%%.

\bibitem{Giusto:2013bda}
S.~Giusto and R.~Russo, ``{Superdescendants of the D1D5 CFT and their dual
  3-charge geometries},'' {\em JHEP} {\bf 03} (2014) 007,
\href{http://arXiv.org/abs/1311.5536}{{\tt 1311.5536}}.
%%CITATION = ARXIV:1311.5536;%%.

\bibitem{Bena:2011dd}
I.~Bena, S.~Giusto, M.~Shigemori, and N.~P. Warner, ``{Supersymmetric Solutions
  in Six Dimensions: A Linear Structure},'' {\em JHEP} {\bf 1203} (2012) 084,
\href{http://arXiv.org/abs/1110.2781}{{\tt 1110.2781}}.
%%CITATION = ARXIV:1110.2781;%%.

\bibitem{Bena:2015bea}
I.~Bena, S.~Giusto, R.~Russo, M.~Shigemori, and N.~P. Warner, ``{Habemus
  Superstratum! A constructive proof of the existence of superstrata},'' {\em
  JHEP} {\bf 05} (2015) 110,
\href{http://arXiv.org/abs/1503.01463}{{\tt 1503.01463}}.
%%CITATION = ARXIV:1503.01463;%%.

\bibitem{Bena:2016ypk}
I.~Bena, S.~Giusto, E.~J. Martinec, R.~Russo, M.~Shigemori, D.~Turton, and
  N.~P. Warner, ``{Smooth horizonless geometries deep inside the black-hole
  regime},'' {\em Phys. Rev. Lett.} {\bf 117} (2016), no.~20, 201601,
\href{http://arXiv.org/abs/1607.03908}{{\tt 1607.03908}}.
%%CITATION = ARXIV:1607.03908;%%.

\bibitem{Bena:2017xbt}
I.~Bena, S.~Giusto, E.~J. Martinec, R.~Russo, M.~Shigemori, D.~Turton, and
  N.~P. Warner, ``{Asymptotically-flat supergravity solutions deep inside the
  black-hole regime},'' {\em JHEP} {\bf 02} (2018) 014,
\href{http://arXiv.org/abs/1711.10474}{{\tt 1711.10474}}.
%%CITATION = ARXIV:1711.10474;%%.

\bibitem{Bombini:2017got}
A.~Bombini and S.~Giusto, ``{Non-extremal superdescendants of the D1D5 CFT},''
  {\em JHEP} {\bf 10} (2017) 023,
\href{http://arXiv.org/abs/1706.09761}{{\tt 1706.09761}}.
%%CITATION = ARXIV:1706.09761;%%.

\bibitem{Bakhshaei:2018vux}
E.~Bakhshaei and A.~Bombini, ``{Three-charge superstrata with internal
  excitations},'' {\em Class. Quant. Grav.} {\bf 36} (2019), no.~5, 055001,
\href{http://arXiv.org/abs/1811.00067}{{\tt 1811.00067}}.
%%CITATION = ARXIV:1811.00067;%%.

\bibitem{Bena:2017fvm}
I.~Bena, P.~Heidmann, and P.~F. Ramirez, ``{A systematic construction of
  microstate geometries with low angular momentum},'' {\em JHEP} {\bf 10}
  (2017) 217,
\href{http://arXiv.org/abs/1709.02812}{{\tt 1709.02812}}.
%%CITATION = ARXIV:1709.02812;%%.

\bibitem{Bena:2018bbd}
I.~Bena, P.~Heidmann, and D.~Turton, ``{AdS$_{2}$ holography: mind the cap},''
  {\em JHEP} {\bf 12} (2018) 028,
\href{http://arXiv.org/abs/1806.02834}{{\tt 1806.02834}}.
%%CITATION = ARXIV:1806.02834;%%.

\bibitem{Heidmann:2018mtx}
P.~Heidmann, ``{Bubbling the NHEK},'' {\em JHEP} {\bf 01} (2019) 108,
\href{http://arXiv.org/abs/1811.08256}{{\tt 1811.08256}}.
%%CITATION = ARXIV:1811.08256;%%.

\bibitem{Walker:2019ntz}
R.~Walker, ``{D1-D5-P superstrata in 5 and 6 dimensions: separable wave
  equations and prepotentials},''
\href{http://arXiv.org/abs/1906.04200}{{\tt 1906.04200}}.
%%CITATION = ARXIV:1906.04200;%%.

\bibitem{Heidmann:2019zws}
P.~Heidmann and N.~P. Warner, ``{Superstratum Symbiosis},''
\href{http://arXiv.org/abs/1903.07631}{{\tt 1903.07631}}.
%%CITATION = ARXIV:1903.07631;%%.

\bibitem{Giusto:2014aba}
S.~Giusto and R.~Russo, ``{Entanglement Entropy and D1-D5 geometries},'' {\em
  Phys. Rev.} {\bf D90} (2014), no.~6, 066004,
\href{http://arXiv.org/abs/1405.6185}{{\tt 1405.6185}}.
%%CITATION = ARXIV:1405.6185;%%.

\bibitem{Giusto:2015dfa}
S.~Giusto, E.~Moscato, and R.~Russo, ``{AdS$_{3}$ holography for 1/4 and 1/8
  BPS geometries},'' {\em JHEP} {\bf 11} (2015) 004,
\href{http://arXiv.org/abs/1507.00945}{{\tt 1507.00945}}.
%%CITATION = ARXIV:1507.00945;%%.

\bibitem{Moscato:2017usq}
E.~Moscato, {\em {Black hole microstates and holography in the D1D5 CFT}}.
\newblock PhD thesis, Queen Mary, U. of London,
2017-09.
\newblock
%%CITATION = INSPIRE-1653282;%%.

\bibitem{Bena:2017upb}
I.~Bena, D.~Turton, R.~Walker, and N.~P. Warner, ``{Integrability and
  Black-Hole Microstate Geometries},'' {\em JHEP} {\bf 11} (2017) 021,
\href{http://arXiv.org/abs/1709.01107}{{\tt 1709.01107}}.
%%CITATION = ARXIV:1709.01107;%%.

\bibitem{Galliani:2016cai}
A.~Galliani, S.~Giusto, E.~Moscato, and R.~Russo, ``{Correlators at large c
  without information loss},'' {\em JHEP} {\bf 09} (2016) 065,
\href{http://arXiv.org/abs/1606.01119}{{\tt 1606.01119}}.
%%CITATION = ARXIV:1606.01119;%%.

\bibitem{Galliani:2017jlg}
A.~Galliani, S.~Giusto, and R.~Russo, ``{Holographic 4-point correlators with
  heavy states},'' {\em JHEP} {\bf 10} (2017) 040,
\href{http://arXiv.org/abs/1705.09250}{{\tt 1705.09250}}.
%%CITATION = ARXIV:1705.09250;%%.

\bibitem{Bombini:2019vnc}
A.~Bombini and A.~Galliani, ``{AdS$_{3}$ four-point functions from $
  \frac{1}{8} $ -BPS states},'' {\em JHEP} {\bf 06} (2019) 044,
\href{http://arXiv.org/abs/1904.02656}{{\tt 1904.02656}}.
%%CITATION = ARXIV:1904.02656;%%.

\bibitem{Tian:2019ash}
J.~Tian, J.~Hou, and B.~Chen, ``{Holographic Correlators on Integrable
  Superstrata},''
\href{http://arXiv.org/abs/1904.04532}{{\tt 1904.04532}}.
%%CITATION = ARXIV:1904.04532;%%.

\bibitem{Bena:2019azk}
I.~Bena, P.~Heidmann, R.~Monten, and N.~P. Warner, ``{Thermal Decay without
  Information Loss in Horizonless Microstate Geometries},''
\href{http://arXiv.org/abs/1905.05194}{{\tt 1905.05194}}.
%%CITATION = ARXIV:1905.05194;%%.

\bibitem{Nishioka:2009un}
T.~Nishioka, S.~Ryu, and T.~Takayanagi, ``{Holographic Entanglement Entropy: An
  Overview},'' {\em J. Phys.} {\bf A42} (2009) 504008,
\href{http://arXiv.org/abs/0905.0932}{{\tt 0905.0932}}.
%%CITATION = ARXIV:0905.0932;%%.

\bibitem{Nishioka:2018khk}
T.~Nishioka, ``{Entanglement entropy: holography and renormalization group},''
  {\em Rev. Mod. Phys.} {\bf 90} (2018), no.~3, 035007,
\href{http://arXiv.org/abs/1801.10352}{{\tt 1801.10352}}.
%%CITATION = ARXIV:1801.10352;%%.

\bibitem{Balasubramanian:2017hgy}
V.~Balasubramanian, A.~Lawrence, A.~Rolph, and S.~Ross, ``{Entanglement shadows
  in LLM geometries},'' {\em JHEP} {\bf 11} (2017) 159,
\href{http://arXiv.org/abs/1704.03448}{{\tt 1704.03448}}.
%%CITATION = ARXIV:1704.03448;%%.

\bibitem{Carlip:1995zj}
S.~Carlip, ``{Lectures on (2+1) dimensional gravity},'' {\em J. Korean Phys.
  Soc.} {\bf 28} (1995) S447--S467,
\href{http://arXiv.org/abs/gr-qc/9503024}{{\tt gr-qc/9503024}}.
%%CITATION = GR-QC/9503024;%%.

\bibitem{Carmi:2016wjl}
D.~Carmi, R.~C. Myers, and P.~Rath, ``{Comments on Holographic Complexity},''
  {\em JHEP} {\bf 03} (2017) 118,
\href{http://arXiv.org/abs/1612.00433}{{\tt 1612.00433}}.
%%CITATION = ARXIV:1612.00433;%%.

\bibitem{Reynolds:2016rvl}
A.~Reynolds and S.~F. Ross, ``{Divergences in Holographic Complexity},'' {\em
  Class. Quant. Grav.} {\bf 34} (2017), no.~10, 105004,
\href{http://arXiv.org/abs/1612.05439}{{\tt 1612.05439}}.
%%CITATION = ARXIV:1612.05439;%%.

\bibitem{Carmi:2017jqz}
D.~Carmi, S.~Chapman, H.~Marrochio, R.~C. Myers, and S.~Sugishita, ``{On the
  Time Dependence of Holographic Complexity},'' {\em JHEP} {\bf 11} (2017) 188,
\href{http://arXiv.org/abs/1709.10184}{{\tt 1709.10184}}.
%%CITATION = ARXIV:1709.10184;%%.

\bibitem{Reynolds:2017jfs}
A.~P. Reynolds and S.~F. Ross, ``{Complexity of the AdS Soliton},'' {\em Class.
  Quant. Grav.} {\bf 35} (2018), no.~9, 095006,
\href{http://arXiv.org/abs/1712.03732}{{\tt 1712.03732}}.
%%CITATION = ARXIV:1712.03732;%%.

\bibitem{Chapman:2018dem}
S.~Chapman, H.~Marrochio, and R.~C. Myers, ``{Holographic complexity in Vaidya
  spacetimes. Part I},'' {\em JHEP} {\bf 06} (2018) 046,
\href{http://arXiv.org/abs/1804.07410}{{\tt 1804.07410}}.
%%CITATION = ARXIV:1804.07410;%%.

\bibitem{Susskind:2018vql}
L.~Susskind, ``{PiTP Lectures on Complexity and Black Holes},''
\newblock 2018.
\newblock
\href{http://arXiv.org/abs/1808.09941}{{\tt 1808.09941}}.
\newblock
%%CITATION = ARXIV:1808.09941;%%.

\bibitem{Susskind:2018pmk}
L.~Susskind, ``{Three Lectures on Complexity and Black Holes},''
\newblock 2018.
\newblock
\href{http://arXiv.org/abs/1810.11563}{{\tt 1810.11563}}.
\newblock
%%CITATION = ARXIV:1810.11563;%%.

\bibitem{Chapman:2018hou}
S.~Chapman, J.~Eisert, L.~Hackl, M.~P. Heller, R.~Jefferson, H.~Marrochio, and
  R.~C. Myers, ``{Complexity and entanglement for thermofield double states},''
  {\em SciPost Phys.} {\bf 6} (2019), no.~3, 034,
\href{http://arXiv.org/abs/1810.05151}{{\tt 1810.05151}}.
%%CITATION = ARXIV:1810.05151;%%.

\bibitem{Balasubramanian:2018hsu}
V.~Balasubramanian, M.~DeCross, A.~Kar, and O.~Parrikar, ``{Binding Complexity
  and Multiparty Entanglement},'' {\em JHEP} {\bf 02} (2019) 069,
\href{http://arXiv.org/abs/1811.04085}{{\tt 1811.04085}}.
%%CITATION = ARXIV:1811.04085;%%.

\bibitem{Goto:2018iay}
K.~Goto, H.~Marrochio, R.~C. Myers, L.~Queimada, and B.~Yoshida, ``{Holographic
  Complexity Equals Which Action?},'' {\em JHEP} {\bf 02} (2019) 160,
\href{http://arXiv.org/abs/1901.00014}{{\tt 1901.00014}}.
%%CITATION = ARXIV:1901.00014;%%.

\bibitem{Bernamonti:2019zyy}
A.~Bernamonti, F.~Galli, J.~Hernandez, R.~C. Myers, S.-M. Ruan, and J.~Sim\'on,
  ``{The First Law of Complexity},''
\href{http://arXiv.org/abs/1903.04511}{{\tt 1903.04511}}.
%%CITATION = ARXIV:1903.04511;%%.

\bibitem{Ross:2019rtu}
S.~F. Ross, ``{Complexity and typical microstates},''
\href{http://arXiv.org/abs/1905.06211}{{\tt 1905.06211}}.
%%CITATION = ARXIV:1905.06211;%%.

\bibitem{Cadoni:2009tk}
M.~Cadoni and M.~Melis, ``{Holographic entanglement entropy of the BTZ black
  hole},'' {\em Found. Phys.} {\bf 40} (2010) 638--657,
\href{http://arXiv.org/abs/0907.1559}{{\tt 0907.1559}}.
%%CITATION = ARXIV:0907.1559;%%.

\end{thebibliography}
\end{document}